\documentclass[10pt]{article}
\usepackage{amsmath}
\usepackage{graphicx}
\usepackage{color}
\usepackage{orcidlink}
\usepackage{hyperref}
\hypersetup{colorlinks=true, linkcolor=blue, citecolor=blue, urlcolor=blue}
\usepackage{caption}
\usepackage{subcaption}
\begin{document}
\baselineskip=14pt

\begin{center}
{\large {\bf Harmonic oscillator system in topologically charged Eddington-inspired Born-Infeld gravity space-time and Wu-Yang magnetic monopole }}
\end{center}

\vspace{0.3cm}

\begin{center}
    {\bf Faizuddin Ahmed\orcidlink{0000-0003-2196-9622}}\footnote{\tt faizuddinahmed15@gmail.com}\\
    \vspace{0.1cm}
    {\it Department of Physics, University of Science \& Technology Meghalaya, Ri-Bhoi, Meghalaya, 793101, India}\\
    \vspace{0.3cm}
    {\bf Abdelmalek Bouzenada\orcidlink{0000-0002-3363-980X}}\footnote{\tt abdelmalekbouzenada@gmail.com }\\
    \vspace{0.1cm}
    {\it  Laboratory of Theoretical and Applied Physics, Echahid Cheikh Larbi Tebessi University, Algeria}\\
\end{center}

\vspace{0.3cm}

\begin{abstract}
We investigate the quantum dynamics of a harmonic oscillator (HO) system within the framework of topologically charged Eddington-inspired Born-Infeld (EiBI) gravity space-time. Additionally, we incorporate the Wu-Yang magnetic monopole (WYMM) into the quantum system and analyze the influences of modified gravity, the topological charge, and WYMM on this HO quantum mechanical system. Using analytical methods, we derive the energy eigenvalues and the corresponding eigenfunctions of the HO. Moreover, we consider a scenario where the EiBI-gravity parameter is absent and introduce an inverse square potential, including the WYMM, into the HO system. Our findings show significant deviations in the behavior of the HO system compared to the traditional quantum mechanical HO model, including alterations in the bound-state spectra and eigenfunctions. These results provide valuable insights into the interplay between quantum mechanical problems and alternative gravitational theories.
\end{abstract}

\vspace{0.1cm}


\vspace{0.1cm}


\section{Introduction}
When Albert Einstein developed General Relativity in 1915 \cite{a1}, it revolutionized our understanding of gravity by characterizing it as the curvature of space-time caused by mass and energy, rather than as a force. This theory provides a comprehensive framework for understanding the large-scale structure of the universe, extending the principles of Special Relativity to include the effects of gravity \cite{a2}. According to General Relativity, massive objects such as planets and stars warp the space-time around them. This curvature affects the motion of other objects, producing the effect we perceive as gravitational attraction. General Relativity has profound implications for our understanding of black holes, the expansion of the universe, and the nature of time and space. It remains a cornerstone of modern physics.

Two of the most fundamental theories in modern physics are General Relativity \cite{a3} and Quantum Mechanics \cite{a4}, each describing unique aspects of the natural world. Quantum Mechanics is founded on the principles of wave-particle duality \cite{a5}, the quantization of energy \cite{a6}, and probabilistic outcomes \cite{a7}, governing the behavior of particles at the smallest scales, including atoms and subatomic particles \cite{a8}. In contrast, General Relativity describes the macroscopic world by using the curvature of space-time to explain gravitational interactions between massive objects. Despite their successes, these two frameworks are fundamentally irreconcilable; General Relativity operates on a deterministic and continuous space-time fabric, while Quantum Mechanics is inherently discrete and probabilistic. The quest to unify these theories into a single coherent framework, often referred to as quantum gravity, remains one of the most profound challenges in modern physics. Advances in this area promise to deepen our understanding of phenomena such as black holes \cite{a9,a10}, the early universe, and the fundamental nature of space and time itself. Various approaches \cite{a11}, including string theory \cite{a12}, loop quantum gravity \cite{a13}, and modifications like Born-Infeld gravity \cite{a14,a15,a16,a17}, aim to bridge this gap, suggesting that the universe at its most fundamental level might be governed by principles that reconcile these two seemingly disparate theories.

An intriguing and intricate field of theoretical physics that combines ideas from General Relativity and Quantum Mechanics is the study of the harmonic oscillator in curved space-time \cite{a18,a19}. Unlike the behavior of a basic harmonic oscillator in flat space-time \cite{a20}, the curvature of space-time itself significantly influences the behavior of such a system in a curved background \cite{a21}. This curvature can be caused by the presence of massive objects or more complex gravitational fields \cite{a22, a22-1}. The wave function and energy spectrum of the oscillator field are altered by this curvature, which also modifies the oscillator's kinetic and potential components \cite{aa22}. By examining the harmonic oscillator in curved space-time, scientists can probe the limits of quantum mechanics in extreme environments, offering insights into potential quantum gravitational effects and advancing the broader goal of developing a unified theory of quantum gravity \cite{a23}. In addition to bridging the gap between General Relativity and Quantum Field Theory, this branch of study has applications in understanding astrophysical processes and particle behavior in the intense gravitational fields of black holes \cite{a24} and neutron stars \cite{a25}.

In General Relativity, numerous modified theories of gravity have been proposed to address issues that cannot be explained by GR alone. Among these theories, Eddington-inspired Born-Infeld (EiBI) gravity \cite{a26, a27} stands out as a fascinating and innovative theoretical framework. EiBI gravity seeks to address some longstanding puzzles in gravitational physics by merging concepts from two influential approaches: the Eddington gravitational theory \cite{a28} and the Born-Infeld theory of electromagnetism. Originally proposed by M. Baños {\it et al.} in the context of non-linear electrodynamics \cite{a29}, the Born-Infeld theory aimed to eliminate the singularities associated with point charges \cite{a31}. When applied to gravity, this idea offers a non-linear modification to General Relativity (GR) \cite{a32}, which can potentially regularize the singularities inherent in standard GR solutions, such as the Big Bang singularity \cite{a33} and black hole singularities \cite{a34}. EiBI gravity posits that the space-time metric and the connection are independent variables \cite{a35}, an idea borrowed from Eddington's 1924 affine gravity theory \cite{a36}. The simplest solution ever obtained in EiBI gravity is the Global Monopole/Wormhole (GM/WH) \cite{pp1,pp2}. This solution interpolates between a modified global monopole \cite{a41} and a wormhole similar to the Ellis–Bronnikov wormhole with topological charge. It is noteworthy that this solution was obtained with a source of matter that does not violate the energy conditions and, impressively, it is as simple as the well-known Ellis-Bronnikov solution \cite{pp3,pp4}.

The implications of the harmonic oscillator (HO) for fundamental physics have drawn considerable interest. Numerous scenarios have been studied for HO solutions, providing insights into the behavior of fields and matter in curved space-times. The HO problem has been examined in various curved space-time backgrounds, such as in a space with a combined linear topological defect-namely, disclination plus a dislocation \cite{ref2}, the topological Aharonov-Bohm effect in a 2D scenario \cite{ref11}, topologically charged Ellis-Bronnikov-type wormholes \cite{ref1}, cosmic string space-time with dislocation under a repulsive inverse square potential and rotational frame effects \cite{ref3}, in the background of a circularly symmetric, static wormhole space-time with cosmic strings \cite{ref5}, Morris-Thorne-type wormhole backgrounds with cosmic strings \cite{ref4}, Born-Infeld space-time background under the Aharonov–Bohm geometric phase \cite{ref12}, a space-time with screw dislocation background under rotational and inverse-square potential effects in the presence of a flux field \cite{ref6}, a space-time with distortions such as a vertical spiral \cite{ref7}, an elastic medium with a spiral dislocation \cite{ref8}, the effect of singular potentials \cite{ref9}, and conical singularities \cite{ref10, RLLV}, influence of spiral dislocation topology on the revival time \cite{ref13}, effects of a screw dislocation and a linear potential \cite{ref14}, non-inertial effects in the presence of a screw dislocation \cite{ref15}.

Our primary objective is to explore the quantum dynamics of a harmonic oscillator system within the context of a topologically charged Eddington-inspired Born-Infeld (EiBI) gravity space-time. Additionally, we incorporate the effects of a Wu-Yang magnetic monopole (WYMM) into the harmonic oscillator system and perform a comprehensive analysis of the eigenvalue solutions. Specifically, we demonstrate that the energy spectrum and wave function of the harmonic oscillator are influenced by the global monopole (GM) and EiBI gravity parameters. In addition, the presence of a WYMM further shifts the energy spectrum and alters the wave function of the harmonic oscillator. Moreover, we examine the harmonic oscillator system within a point-like defect geometry, considering the influences of a WYMM and a repulsive inverse square potential. We illustrate how these parameters affect the energy spectrum and the radial wave function. It is shown that the eigenvalue solutions of the harmonic oscillator system gets modification by these parameters. The point-like global monopole space-time in spherical coordinates $(t, r, \theta, \varphi)$ is described by the following line-element \cite{a41}
\begin{equation}
    ds^2=-dt^2+\frac{dr^2}{\alpha^2}+r^2\,(d\theta^2+\sin^2 \theta d\varphi^2),\label{a}
\end{equation}
where $\alpha=(1-8\,\pi\,G\,\zeta^2)$ is the global monopole (GM) parameter and $\zeta$ is the energy scale of the symmetry breaking.

This paper is organized as: In {\bf Section 2}, we investigate the harmonic oscillator problem within the Eddington-inspired Born-Infeld (EiBI) gravity in global monopole (GM) space-time. In {\bf Section 3}, we consider the harmonic oscillator problem without EiBI gravity in global monopole space-time but in the presence of a Wu-Yang magnetic monopole (WYMM) and an external inverse square potential, and analyze the results. In {\bf Section 4}, we study the harmonic oscillator problem in GM space-time within EiBI gravity and in the presence of a WYMM. In {\bf Section 5}, we present the conclusions. Throughout the article, we choose the system of units, where $c=1=\hbar=G$.

\section{Harmonic oscillator in topologically charged EiBI-gravity space-time}

In this section, we study the harmonic oscillator problem within the topologically charged Eddington-inspired Born-Infeld (EiBI) gravity space-time and analyze the results. 

Therefore, we begin this section by writing the line-element describing this topologically charged EiBI-gravity space-time in spherical coordinates $(t, r, \theta, \varphi)$ given by \cite{c6, c7}
\begin{equation}
    ds^2=-dt^2+\frac{dr^2}{\alpha^2\,\left(1+\frac{\kappa}{r^2}\right)}+r^2\,d\Omega^2,\label{a1}
\end{equation}
where $d\Omega^2=(d\theta^2+\sin^2 \theta d\varphi^2)$, and $\kappa$ is the Eddington parameter that controls the non-linearity of EiBI-gravity. When the Eddington parameter is switched off, {\it i.e.,} $\kappa=0$, this metric (\ref{a1}) reduces to GM one, Eq. (\ref{a}) above. Therefore, the line-element (\ref{a1}) is sometimes referred to as the modified GM or topologically charged EiBI-gravity space-time. One of the most intriguing features of EiBI- gravity is its ability to avoid cosmological singularities, resulting in entirely singularity-free states \cite{c1, c2}. When $\alpha=1$ and $\kappa=-b^2<0$, where $b=const$, it corresponds to a Morris-Thorne-type wormhole space-time \cite{ref4, c3, c4, c8, c9}. Furthermore, when $\kappa <0$, it describes a topologically charged wormhole space-time \cite{c10}.

Transforming to a new variable via $r=\sqrt{\rho^2-\kappa}$ into the above EiBI-gravity space-time (\ref{a1}) results the following metric given by
\begin{equation}
    ds^2=-dt^2+\frac{d\rho^2}{\alpha^2}+(\rho^2-\kappa)\,d\Omega^2.\label{aa1}
\end{equation}
It is worthwhile mentioning that our primary interest lies in investigating the harmonic oscillator problem in the space-time described by (\ref{aa1}). By expressing the above space-time in the form $ds^2=-dt^2+g_{ij}\,dx^i\,dx^j$, we obtain the corresponding spatial metric tensor $g_{ij}$ and its contravariant form as follows:
\begin{equation}
    g_{ij}=\begin{pmatrix}
        \frac{1}{\alpha^2} & 0 & 0\\
        0 & (\rho^2-\kappa) & 0\\
        0 & 0 & (\rho^2-\kappa)\,\sin^2 \theta
    \end{pmatrix},\quad 
    g^{ij}=\begin{pmatrix}
        \alpha^2 & 0 & 0\\
        0 & \frac{1}{(\rho^2-\kappa)} & 0\\
        0 & 0 & \frac{1}{(\rho^2-\kappa)\,\sin^2 \theta}
    \end{pmatrix}.\label{a2}
\end{equation}
Its determinant is given by
\begin{equation}
    det (g_{ij})=g=\frac{(\rho^2-\kappa)^2\,\sin^2 \theta}{\alpha^2}.\label{a3}
\end{equation}

Now, we introduce the harmonic oscillator system in the presence of an external potential $V({\bf r})$ described by the following non-relativistic wave equation
\begin{equation}
    \left[-\frac{1}{2\,M}\,\frac{1}{\sqrt{g}}\,\partial_{i}\,\left (g^{ij}\,\sqrt{g}\,\partial_{j} \right)+\frac{1}{2}\,M\,\omega^2\,\rho^2+V({\bf r})\right]\,\Psi ({\bf r})=E\,\Psi ({\bf r}),\label{a4}
\end{equation}
where $M$ and $E$ are the particles' mass and energy, respectively, $\omega$ is the frequency of the harmonic oscillator system, $g$ is the determinant of the spatial metric tensor $g_{ij}$, and $\Psi({\bf r})$ is the wave function.

Form the spatial metric tensor $g_{ij}$ given in Eq. (\ref{a2}), it is clear that this metric tensor depends on $(\rho, \theta)$ while being independent of the coordinate $(\varphi)$. Let us consider the wave function ansatz $\Psi({\bf r})$ in the following form:  
\begin{equation}
    \Psi ({\bf r})=R(\rho)\,Y_{\ell, m} (\theta, \varphi),\label{a5}
\end{equation}
where $R(\rho)$ is the radial wave function, $Y_{\ell,m} (\theta, \varphi)=\Phi_{m}(\varphi)\,\Theta_{\ell,m}(\theta)$ is the spherical harmonics, and $\ell, m$, respectively are the angular and azimuthal quantum numbers.

Thereby expressing the wave equation (\ref{a4}) using (\ref{a2}) and the wave function ansatz (\ref{a5}), we obtain the following differential equation given by 
\begin{eqnarray}
    \Bigg[\alpha^2\left\{\frac{d^2}{d\rho^2}+\frac{2\,\rho}{(\rho^2-\kappa)}\,\frac{d}{d\rho}\right\}+2\,M\,(E-V(\rho))-M^2\,\omega^2\,\rho^2-\frac{\ell\,(\ell+1)}{(\rho^2-\kappa)}\Bigg]\,R(\rho)=0,\label{a6}
\end{eqnarray}
where we have used the following angular relations
\begin{equation}
    -\Bigg[\frac{1}{\sin \theta}\,\frac{d}{d\theta}\,\left(\sin \theta\,\frac{d}{d\theta}\right)+\frac{1}{\sin^2 \theta}\,\frac{d^2}{d\varphi^2}\Bigg]\,Y_{\ell,m}(\theta, \varphi)=\ell\,(\ell+1)\,Y_{\ell, m}(\theta, \varphi),\label{a7}
\end{equation}
where $\ell$ is the angular quantum number which is related with the magnetic quantum number $m$ as $\ell=(|m|+n)$ with $n=0,1,2,...$. The solutions of the angular equations are well-known and given in many quantum mechanics textbooks.

Now, we try to solve the above radial equation (\ref{a6}) and obtain analytical solutions of the quantum system under investigation. Considering zero external potential, $V(\rho)=0$, the radial wave equation (\ref{a6}) can be written as:
\begin{eqnarray}
    R''(\rho)+\frac{2\,\rho}{\rho^2-\kappa}\,R'(\rho)+\Bigg[\Lambda-\Omega^2\,\rho^2-\frac{\tau^2}{\rho^2-\kappa}\Bigg]\,R(\rho)=0\,,\label{a8}
\end{eqnarray}
where we have defined
\begin{equation}
    \Lambda=\frac{2\,M\,E}{\alpha^2},\quad\quad \Omega=\frac{M\,\omega}{\alpha},\quad\quad \tau^2=\frac{\ell\,(\ell+1)}{\alpha^2}\,.\label{aa8}
\end{equation}

Transforming to a new function via 
\begin{equation}
    R(\rho)=\exp \left(-\frac{1}{2}\,\Omega\,\rho^2\right)\,\psi(\rho)\label{a9}
\end{equation}
into the radial Eq. (\ref{a8}) results the following differential equation form:
\begin{eqnarray}
    \psi''(\rho)+\Bigg[\frac{2\,\rho}{\rho^2-\kappa}-2\,\Omega\,\rho\Bigg]\,\psi'(\rho)+\Bigg[\Delta-\frac{2\,\Omega\,\rho^2}{\rho^2-\kappa}-\frac{\tau^2}{\rho^2-\kappa}\Bigg]\,\psi(\rho)=0,\label{a10}
\end{eqnarray}
where we set the parameter
\begin{equation}
    \Delta=\Lambda-\Omega=\frac{2\,M\,E}{\alpha^2}-\Omega.\label{a11}
\end{equation}

Finally transforming to a new variable via $\rho^2=s\,\kappa$ into the equation (\ref{a10}) results the following second-order differential equation form given by
\begin{eqnarray}
    \psi''(s)+\left[-\Omega\,\kappa+\frac{1/2}{s}+\frac{1}{s-1}  \right]\,\psi'(s)+\Bigg[\frac{\left(\Delta\,\kappa+\tau^2\right)/4}{s}+\frac{-\left(2\,\Omega\,\kappa+\tau^2\right)/4}{s-1} \Bigg]\,\psi(s)=0.\label{a12}
\end{eqnarray}

Equation (\ref{a12}) is the confluent Heun differential equation form \cite{AR,SYS}. Comparing Eq. (\ref{a12}) with the Eq. (A1) in appendix, we obtain
\begin{equation}
    \Xi=-\Omega\,\kappa,\quad\quad \beta=-1/2,\quad\quad \gamma=0,\quad\quad \mu=\left(\Delta\,\kappa+\tau^2\right)/4,\quad\quad \nu=-\left(2\,\Omega\,\kappa+\tau^2\right)/4.\label{a13}
\end{equation}

Thus, $\psi (s)$ in Eq. (\ref{a12}) is the confluent Heun function given by
\begin{equation}
    \psi(s)=H_{c}\left(-\Omega\,\kappa\,, -\frac{1}{2}\,, 0\,, \frac{M\,E\,\kappa}{2\,\alpha^2}\,, \frac{1}{4}-\frac{(2\,M\,E\,\kappa+\ell\,(\ell+1))}{4\,\alpha^2}\,;s\right).\label{a14}
\end{equation}

In order to solve Eq. (\ref{a12}), we consider a power series expansion given in the form $H_{c}(s)=\sum^{\infty}_{i=0}\,d_{i}\,s^{i}$ \cite{GBA}. Thereby, substituting this power series into the Eq. (\ref{a12}), we obtain following recurrence relation
\begin{eqnarray}
    d_{i+2}=\frac{1}{(i+2)(2\,i+3)}\Big[\Big\{(i+1)(2\,i+3+2\,\Omega\,\kappa)-2\,\mu\Big\}\,d_{i+1}+2\,(-\Omega\,\kappa\,i+\mu+\nu)\,d_{i}\Big] \label{a15}
\end{eqnarray}
with a few coefficients given by
\begin{eqnarray}
    d_1&=&-2\,\mu\,d_0,\nonumber\\
    d_2&=&\frac{1}{6}\Big[\Big\{(3+2\,\Omega\,\kappa)-2\,\mu\Big\}\,d_{1}+2\,(\mu+\nu)\,d_{0}\Big].\label{a16}
\end{eqnarray}

To find solutions for the quantum system under investigation, it is necessary to truncate the power series $H_{c}(s)$ to a finite-degree polynomial of order $n$ so that the radial wave function (\ref{a9}) remains regular everywhere. This truncation is achieved by ensuring the coefficient $d_{n+1}=0$ in the recurrence relation (\ref{a15}), where $i=(n-1)$. Consequently, the recurrence relation Eq. (\ref{a15}) under this condition can be rewritten as follows:
\begin{eqnarray}
    d_n=-\frac{2\,\Big[-\Omega\,\kappa\,(n-1)+\mu+\nu\Big]}{\Big[n\,(2\,n+1+2\,\Omega\,\kappa)-2\,\mu\Big]}\,d_{n-1},\label{a17}
\end{eqnarray}
where $n=1,2,3,...$ represent the radial modes of the system. Analyzing Eq. (\ref{a17}) requires imposing specific values for the radial mode $n$, due to the dependence of the coefficients on $d_n$ and $d_{n-1}$. Let us consider the radial mode $n=1$, which represents the lowest state of the quantum system. Substituting $n=1$ into Eq. (\ref{a17}), we obtain the following coefficient 
\begin{eqnarray}
    d_1=-\frac{2\,(\mu+\nu)}{\Big[3+2\,\Omega\,\kappa-2\,\mu\Big]}\,\,d_{0}.\label{a18}
\end{eqnarray}

Thus, by combining Eqs. (\ref{a16}) and (\ref{a18}) gives us the following expression of the energy eigenvalues given by
\begin{equation}
    E_{1,\ell}=\frac{1}{2\,M}\Bigg[3\,M\,\omega\,\alpha-\frac{\ell\,(\ell+1)}{\kappa}+\frac{2\,\alpha^2}{\kappa}\pm\,\frac{\alpha^2}{\kappa}\,\sqrt{\frac{2\,\ell\,(\ell+1)}{\alpha^2}+4+\frac{4\,M\,\omega\,\kappa}{\alpha}\,\left(\frac{M\,\omega\,\kappa}{\alpha}+4\right)}\Bigg].\label{a19}
\end{equation}

Equation (\ref{a19}) represents the allowed energy values for the lowest energy state of the harmonic oscillator system in topologically charged EiBI-gravity space-time. Notably, the energy spectrum of the harmonic oscillator cannot be determined by a closed expression. On the contrary, it is only possible to determine the allowed energy values for the quantum system by separately imposing values of $n$ into the recurrence relation Eq. (\ref{a17}).

Furthermore, unlike in Ref. \cite{EAFB}, the lowest energy state of the system is not defined by the radial mode $n=0$ but by the radial mode $n=1$. Additionally, it is evident that the allowed energy values for the lowest energy state of the harmonic oscillator depends on the topological and Eddington parameters, $(\alpha, \kappa)$. Moreover, the ground state energy value is influenced by the oscillator frequency $\omega$ and the angular quantum number $\ell$.

The eigenfunction corresponding to the ground state of the energy spectrum given by Eq. (\ref{a19}) is described by the first term of the polynomial solution to the confluent Heun equation, Eq. (\ref{a14}). This eigenfunction is defined as $\psi_{1,\ell}(s)=(d_0+d_1\,s)$. Here $d_0$ and $d_1$ are the coefficients determined by the boundary conditions and the specifics of the quantum system under consideration. This form represents the simplest non-trivial solution, encapsulating the fundamental characteristics of the ground state within the given gravitational framework. Therefore, the ground state wave function is given by
\begin{equation}
    R_{1,\ell} (\rho)=\exp\left(-\frac{M\,\omega}{2\,\alpha}\,\rho^2\right)\,\Bigg[1-\rho^2\,\left\{\frac{1}{\kappa}+\frac{M\,\omega}{\alpha}\pm\frac{1}{2\,\kappa}\,\sqrt{\frac{2\,\ell\,(\ell+1)}{\alpha^2}+4+\frac{4\,M\,\omega\,\kappa}{\alpha}\,\left(\frac{M\,\omega\,\kappa}{\alpha}+4\right)}\right\}\Bigg]\,d_0.\label{a20}
\end{equation}

We observe that the ground state wave function of the harmonic oscillator depends on the topological and Eddington parameters,  $(\alpha, \kappa)$. Additionally, it is influenced by the oscillator frequency $\omega$ and the angular quantum number $\ell$. This dependence reflects the complex interplay between the gravitational effects and the intrinsic characteristics of the harmonic oscillator system in the modified space-time described by the given metric (\ref{aa1}).



Now, we discuss a special case corresponds to zero oscillator frequency, $\omega=0$. Thus, by making $\omega=0$ in the energy spectrum Eq. (\ref{a19}), we obtain the following expression of the allowed values of the lowest energy state given by
\begin{equation}
    E_{1,\ell}=\frac{1}{2\,M}\Bigg[-\frac{\ell\,(\ell+1)}{\kappa}+\frac{2\,\alpha^2}{\kappa}\pm\,\frac{\alpha^2}{\kappa}\,\sqrt{\frac{2\,\ell\,(\ell+1)}{\alpha^2}+4}\Bigg].\label{a21}
\end{equation}
That is, the allowed energy values for the lowest energy state of a non-relativistic quantum particle in topologically charged EiBI-gravity space-time are discrete. This indicates that, even in the absence of harmonic interaction, the non-relativistic quantum particle continues to exhibit discrete energy levels. This characteristic confinement arises from the gravitational effects produced by the topological space-time within the framework of EiBI-gravity.

The corresponding eigenfunction in this special case using Eq. (\ref{a21}) can be expressed as
\begin{equation}
    R_{1,\ell} (\rho)=\Bigg[1-\rho^2\,\left\{\frac{1}{\kappa}\pm\frac{1}{2\,\kappa}\,\sqrt{\frac{2\,\ell\,(\ell+1)}{\alpha^2}+4}\right\}\Bigg]\,d_0.\label{a22}
\end{equation}



\begin{center}
\begin{figure}[ht!]
\subfloat[]{\centering{}\includegraphics[scale=0.32]{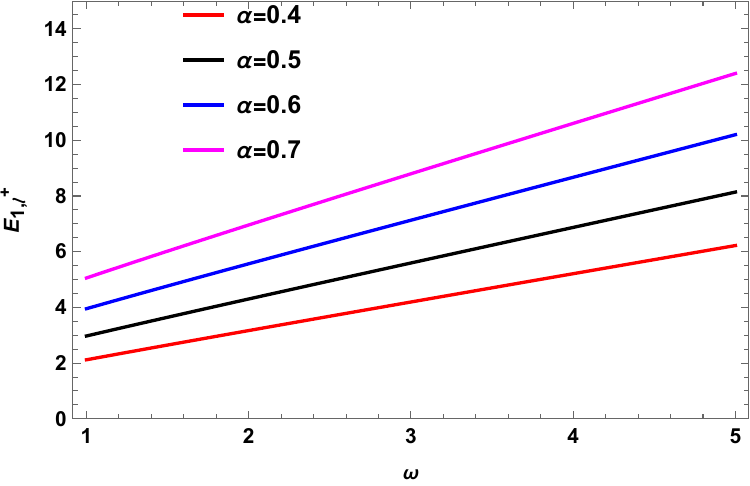}}\quad
\subfloat[]{\centering{}\includegraphics[scale=0.32]{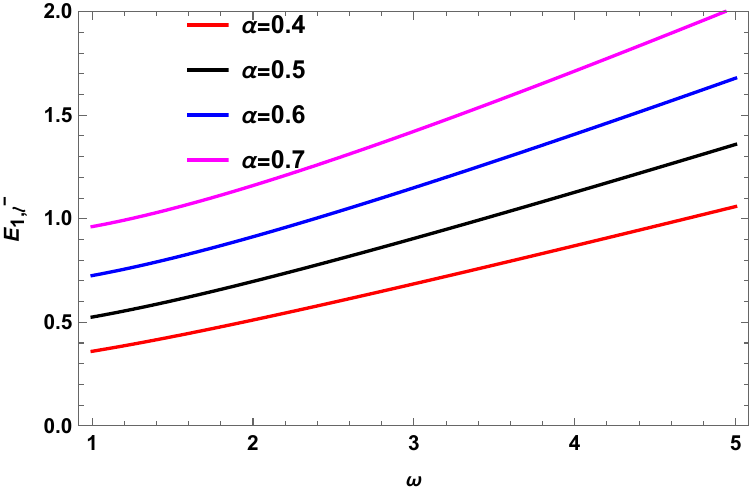}}\quad
\subfloat[]{\centering{}\includegraphics[scale=0.32]{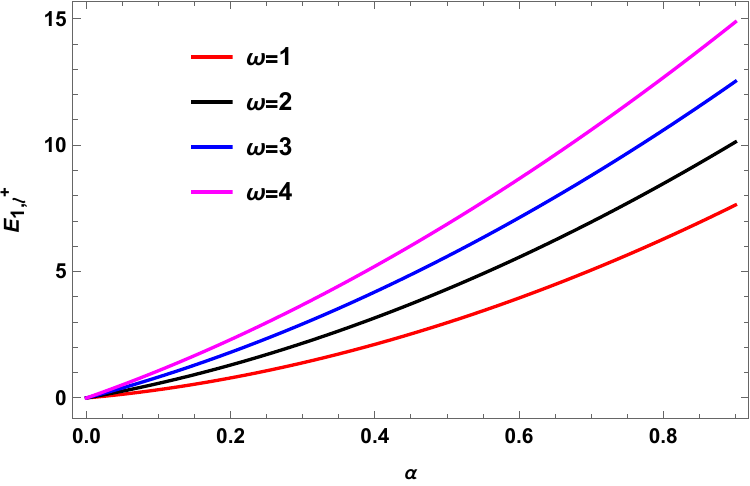}}\quad
\subfloat[]{\centering{}\includegraphics[scale=0.32]{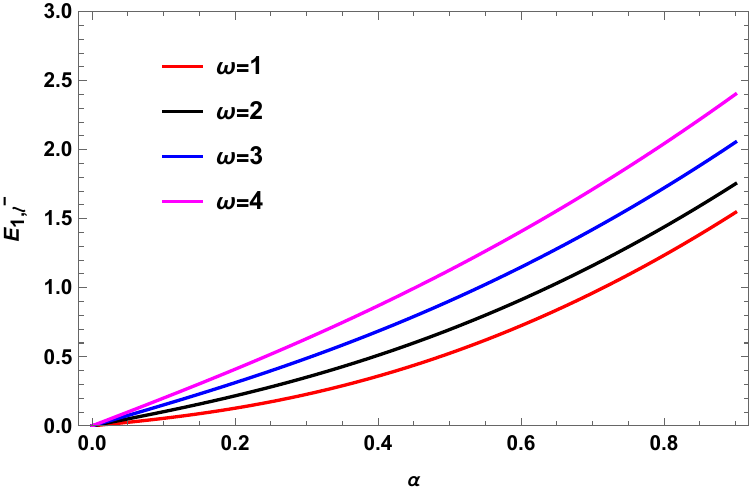}}
\centering{}\caption{The energy level $E_{1,\ell}$ of Eq. (\ref{a19}) for $\ell=0$-state. Here, $E^{+}$ indicates positive sign within the energy expression and $E^{-}$ that for negative sign. Here $M=1$, $\kappa=0.5$.}\label{fig1}
\hfill\\
\subfloat[]{\centering{}\includegraphics[scale=0.32]{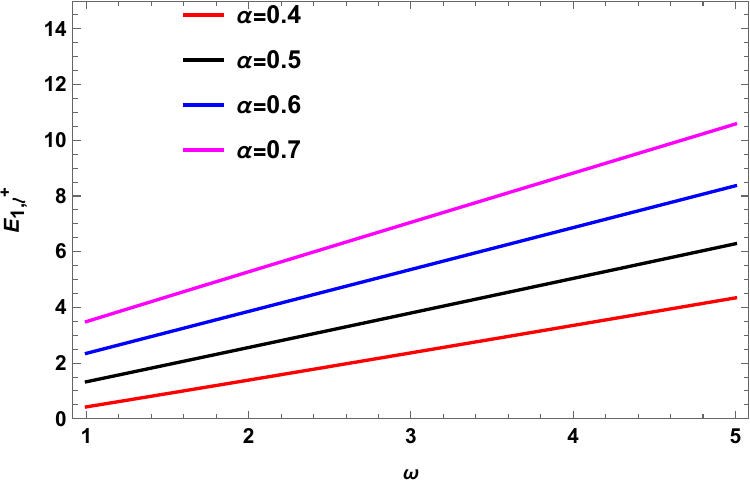}}\quad
\subfloat[]{\centering{}\includegraphics[scale=0.32]{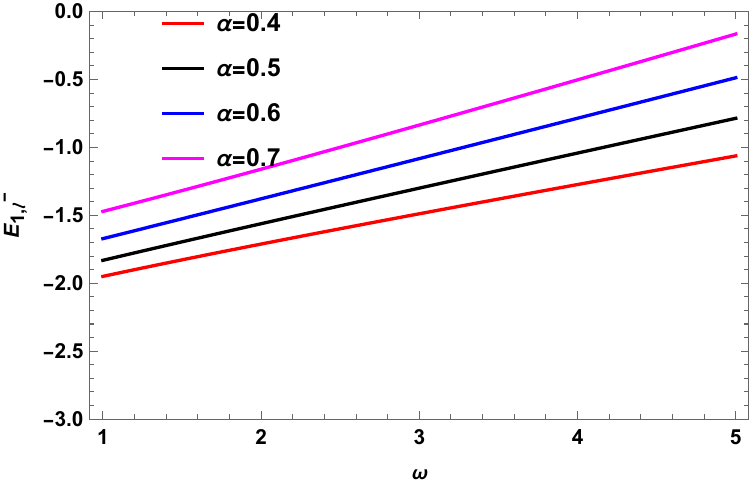}}\quad
\subfloat[]{\centering{}\includegraphics[scale=0.32]{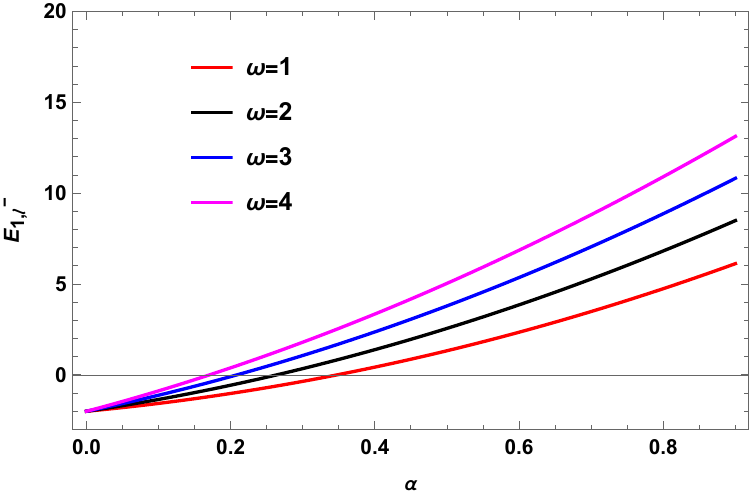}}\quad
\subfloat[]{\centering{}\includegraphics[scale=0.32]{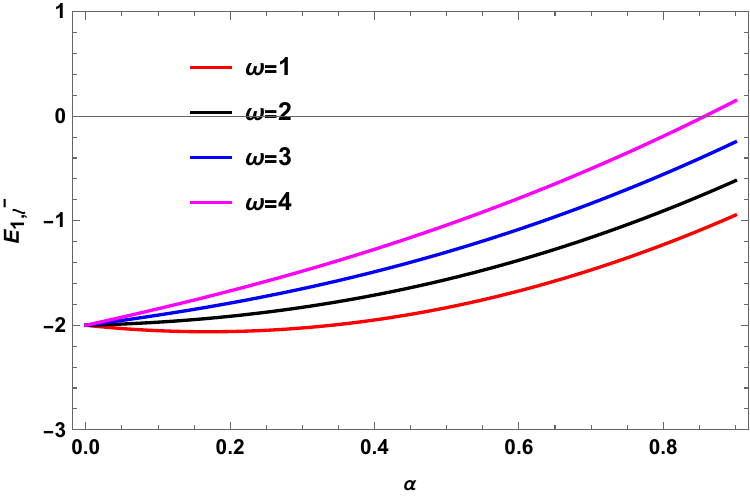}}
\centering{}\caption{The energy level $E_{1,\ell}$ of Eq. (\ref{a19}) for $\ell=1$-state. Here, $E^{+}$ indicates positive sign within the energy expression and $E^{-}$ that for negative sign. Here $M=1$, $\kappa=0.5$.}\label{fig2}
\end{figure}
\par\end{center}

\begin{center}
\begin{figure}[ht!]
\subfloat[$\omega=1$]{\centering{}\includegraphics[scale=0.32]{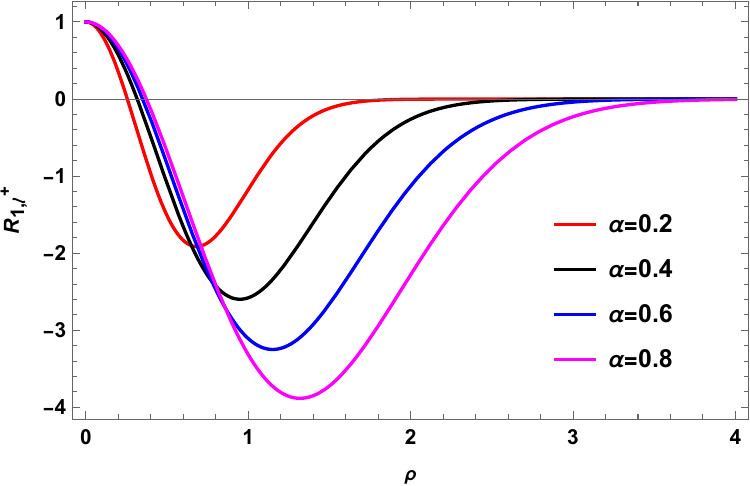}}\quad
\subfloat[$\omega=1$]{\centering{}\includegraphics[scale=0.32]{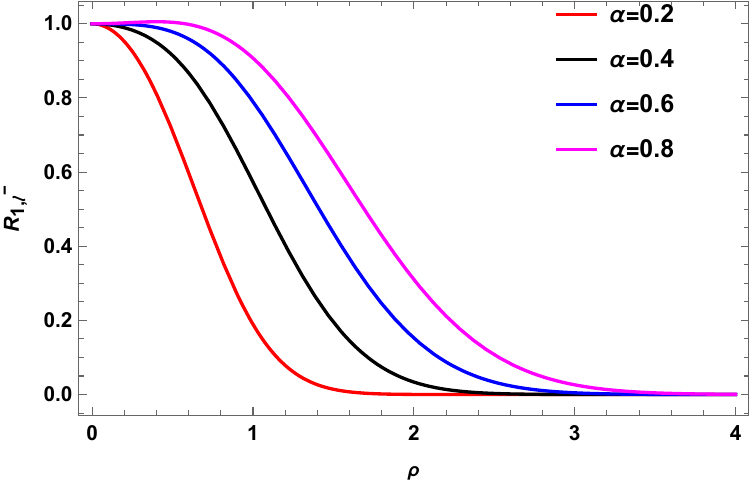}}\quad
\subfloat[$\alpha=0.5$]{\centering{}\includegraphics[scale=0.32]{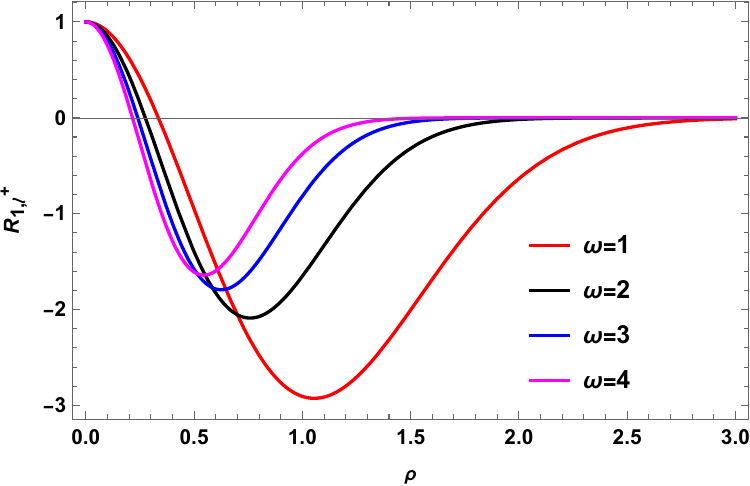}}\quad
\subfloat[$\alpha=0.5$]{\centering{}\includegraphics[scale=0.32]{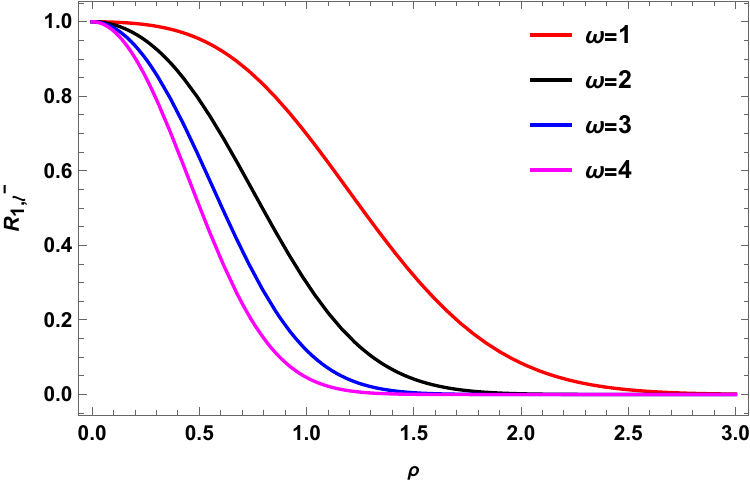}}
\centering{}\caption{The wave function $R_{1,\ell}$ of Eq. (\ref{a20}) for $\ell=0$-state. Here, $R^{+}$ indicates positive sign within the expression and $R^{-}$ that for negative sign. Here $M=1$, $\kappa=0.5$.}\label{fig-wave-function1}
\end{figure}
\par\end{center}

\begin{center}
\begin{figure}[ht!]
\subfloat[$\omega=1$]{\centering{}\includegraphics[scale=0.32]{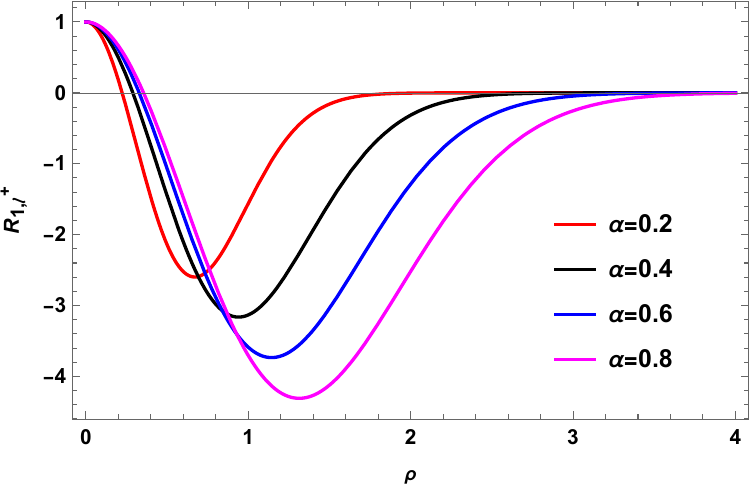}}\quad
\subfloat[$\omega=1$]{\centering{}\includegraphics[scale=0.32]{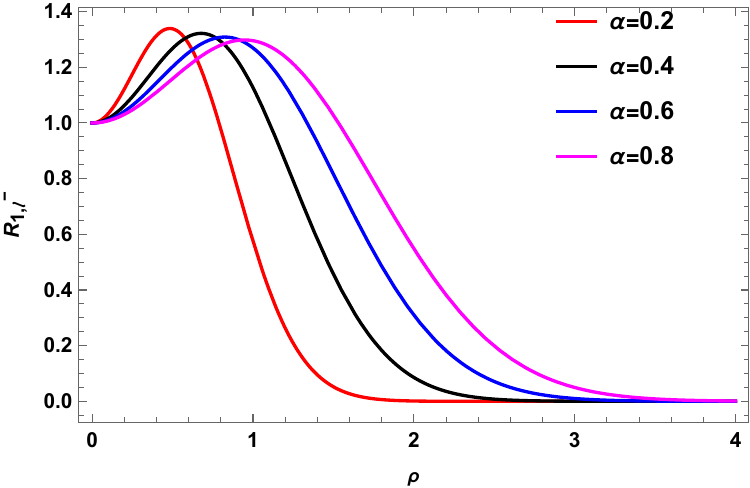}}\quad
\subfloat[$\alpha=0.5$]{\centering{}\includegraphics[scale=0.32]{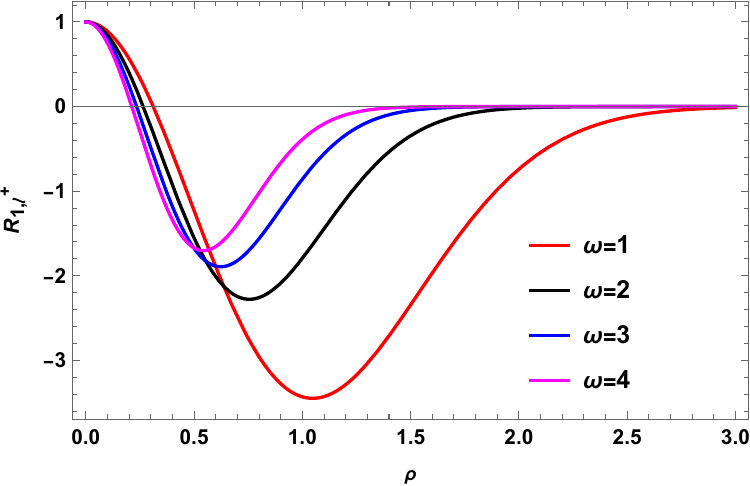}}\quad
\subfloat[$\alpha=0.5$]{\centering{}\includegraphics[scale=0.32]{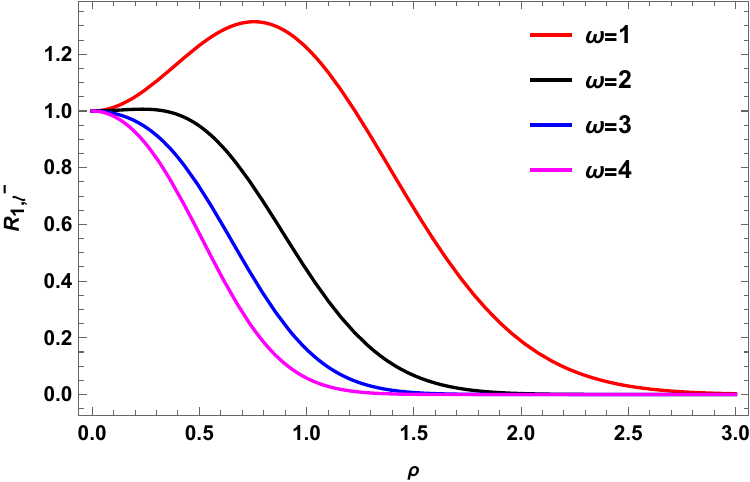}}
\centering{}\caption{The wave function $R_{1,\ell}$ of Eq. (\ref{a20}) for $\ell=1$-state. Here, $R^{+}$ indicates positive sign within the expression and $R^{-}$ that for negative sign.}\label{fig-wave-function2}
\end{figure}
\par\end{center}

We have generated several graphs illustrating the behavior of the ground state energy spectrum, given by Eq. (\ref{a19}), for the $\ell=0$ and $\ell=1$ states. In Figures 1(a) and 1(b) for the $\ell=0$ states, we observe that the energy level linearly increases with the oscillator frequency $\omega$, and this increase shifts upward as the topology parameter $\alpha$ increases. In Figures 1(c) and 1(d), the energy level gradually increases with the topology parameter $\alpha$, and this level shifts upward as the oscillator frequency $\omega$ increases. A similar pattern is observed for the $\ell=1$ state in Figure 2. Additionally, Figures 3 and 4 depict the behavior of the ground state wave function (\ref{a20}) for these two states, $\ell=0$ and $\ell=1$, respectively.

\section{Harmonic oscillator system in topologically charged space-time and a WYMM without EiBI-gravity }

In this part, we study the harmonic oscillator system in topologically charged space-time without Eddington parameter, $\kappa=0$ in the presence of a Wu-Yang magnetic monopole (WYMM). In this WYMM environment, the electromagnetic vector potential $\vec{A}$ is introduced into the non-relativistic wave equation through minimal substitution $\mathcal{D}_{i} \equiv (\partial_{i}-i\,e\,A_{i})$. Consequently, the wave equation describing the harmonic oscillator system in curved space is given by:
\begin{equation}
    \left[-\frac{1}{2\,M}\,\frac{1}{\sqrt{g}}\,\mathcal{D}_{i}\,\left (g^{ij}\,\sqrt{g}\,\mathcal{D}_{j} \right)+\frac{1}{2}\,M\,\omega^2\,\rho^2+V({\bf r})\right]\,\Psi ({\bf r})=E\,\Psi ({\bf r}),\label{b1}
\end{equation}
where $e$ is the electric charge and $E$ is the particle's energy.

In Ref. \cite{WY1}, the most elegant formalism to describe an Abelian point-like magnetic monopole were presented. It is shown that the electromagnetic vector potential can be described by a singularity-free expression. The electromagnetic vector potential, defined by $A_{\mu}=(-A_0, \vec{A})$, is expressed in two overlapping regions, $R_a$ and $R_b$, which together cover the entire space. Using spherical coordinate system, with the monopole located at the origin, the vector potential in these regions is given by:
\begin{align}
    R_a&: 0\leq \theta < \frac{\pi}{2}+\delta,  & r>0,\quad 0 \leq \phi < 2\,\pi,\nonumber\\
    R_b&: \frac{\pi}{2}-\delta < \theta \leq \pi,  & r>0, \quad 0 \leq \phi < 2\,\pi,\nonumber\\
    R_{ab}&: \frac{\pi}{2}-\delta < \theta < \frac{\pi}{2}+\delta,  & r>0, \quad 0 \leq \phi < 2\,\pi,\label{WY1}
\end{align}
where $0 < \delta \leq \pi/2$ and $R_{ab}$ being the overlapping region.

The only non vanishing components of vector potential are
\begin{align}
    (A_{\phi})_a&=g\,(1-\cos \theta),\nonumber\\
    (A_{\phi})_b&=-g\,(1+\cos \theta),\label{WY2}
\end{align}
$g$ being the monopole strength. In the overlapping region the non vanishing components are related by a gauge transformation
\begin{equation}
    (A_{\phi})_a=(A_{\phi})_a+\frac{i}{e}\,S\,\partial_{\phi}\,S^{-1},\quad S=e^{2\,i\,\sigma\,\phi}.\label{WY3}
\end{equation}
In order to have single valued gauge transformation, we must have $\sigma=e\,g$.  This is known as the Dirac quantization condition.

According to the analysis by Wu and Yang in Ref. \cite{WY2}, the solutions of the non-relativistic wave equation in the presence of such a vector potential are not ordinary functions but rather sections. Specifically, the solution assumes the values $\Psi_a$ and $\Psi_b$ in $R_a$ and $R_b$, respectively, and satisfies the gauge transformation:
\begin{equation}
    \Psi_a=S\,\Psi_b\,.\label{WY4}
\end{equation}
This gauge transformation ensures the consistency of the wave function in the overlap region $R_{ab}$ between $R_a$ and $R_b$, reflecting the underlying topological nature of the magnetic monopole. Thus, the wave function $\Psi$ in the presence of the Wu-Yang magnetic monopole is described by sections that transform appropriately under the given gauge transformation.

Therefore, explicitly writing the harmonic oscillator equation (\ref{b1}) in topologically charged space-time (\ref{a}) given by
\begin{eqnarray}
    &&\frac{r^2}{R(r)}\,\Bigg[\alpha^2\left(\frac{d^2}{dr^2}+\frac{2}{r}\,\frac{d}{dr}\right)\,R(r)+\Big\{2\,M\,(E-V(r))-M^2\,\omega^2\,r^2\Big\}\,R(r)\Bigg]\nonumber\\
    &&=-\frac{1}{Y_{\ell^{\prime}, m^{\prime}}(\theta, \varphi)}\,\Bigg[\frac{1}{\sin \theta}\,\frac{d}{d\theta}\,\left(\sin \theta\,\frac{d\,Y_{\ell^{\prime}, m^{\prime}}(\theta, \varphi)}{d\theta}\right)+\frac{1}{\sin^2 \theta}\,\left(\frac{d}{d\varphi}-A_{\varphi}\right)^2\,\,Y_{\ell^{\prime}, m^{\prime}}(\theta, \varphi)\Bigg].\label{b3}
\end{eqnarray}
Here $Y_{\ell^{\prime},m^{\prime}}$ is called the conical monopole harmonics and $\ell^{\prime}, m^{\prime}$ are the modified quantum numbers due to the presence of WYMM.

Writing the separation constant $\lambda^{\prime}$, we can write the angular equations as follows:
\begin{eqnarray}
    \Bigg[\frac{1}{\sin \theta}\,\frac{d}{d\theta}\,\left(\sin \theta\,\frac{d}{d\theta}\right)+\frac{1}{\sin^2 \theta}\,\left(\frac{d}{d\varphi}-A_{\phi}\right)^2\Bigg]\,Y_{\ell^{\prime}, m^{\prime}}(\theta, \varphi)=-\lambda^{\prime}\,Y_{\ell^{\prime}, m^{\prime}}(\theta, \varphi),\label{b4}
\end{eqnarray}
where the eigenvalue $\lambda^{\prime}$ is given by \cite{ALCO1,ALCO2,ALCO3}
\begin{equation}
    \lambda^{\prime}=\ell^{\prime}\,(\ell^{\prime}+1),\quad \ell^{\prime}=-\frac{1}{2}+\sqrt{\frac{1}{4}+\ell\,(\ell+1)-\sigma^2},\quad \sigma=e\,g,\quad m^{\prime}=m+s\,q\quad (s=\pm\,1).\label{b5}
\end{equation}
Notice that for zero electromagnetic potential $A_{\phi}=0$, there is no WYMM, and hence, one will get back the eigenvalue of the angular equations $\lambda=\ell\,(\ell+1)$, where $\ell=(|m|+n)\quad (n=0,1,2...)$ which is given in quantum mechanics text books. For details the above angular eigenvalue solution, readers are suggested to see reference \cite{ALCO1,ALCO2,ALCO3}.

With these, we can write the radial equation for the harmonic oscillator system in topologically charged space-time and in a WYMM without EiBI-gravity using Eq. (\ref{b3}) as follows:
\begin{eqnarray}
    R^{\prime\prime}(r)+\frac{2}{r}\,R^{\prime} (r)+\frac{1}{\alpha^2}\,\Bigg[2\,M\,(E-V(r))-M^2\,\omega^2\,r^2-\frac{\lambda^{\prime}}{r^2} \Bigg]\,R(r)=0.\label{b6}
\end{eqnarray}

Below, we solve the radial equation (\ref{b6}) for two scenarios: the first scenario without any external potential, $V({\bf r})=0$, and the second scenario with an external potential of the form, $V({\bf r})=V(r) \propto \frac{1}{r^2}$. 

\subsection{Zero external potential, V(r)=0.}

In this part, we first consider zero external potential, $V(r)=0$ into the harmonic oscillator system under investigations. Therefore, the radial equation (\ref{b6}) becomes
\begin{eqnarray}
    R^{\prime\prime}(r)+\frac{2}{r}\,R^{\prime} (r)+\frac{1}{\alpha^2}\,\Bigg[2\,M\,E-M^2\,\omega^2\,r^2-\frac{\lambda^{\prime}}{r^2} \Bigg]\,R(r)=0.\label{b7}
\end{eqnarray}

Transforming first $R(r)=\frac{\psi(r)}{\sqrt{r}}$, and then to a new variable $s=\Omega\,r^2$ into the Eq. (\ref{b7}), we obtain the following differential equation 
\begin{equation}
    \psi''(s)+\frac{1}{s}\,\psi'(s)+\Bigg[-\frac{j^2/4}{s^2}-\frac{1}{4}+\frac{\Lambda}{4\,\Omega\,s}  \Bigg]\,\psi(s)=0,\label{b8}
\end{equation}
where
\begin{equation}
    \Lambda=\frac{2\,M\,E}{\alpha^2},\quad j=\sqrt{\frac{\lambda^{\prime}}{\alpha^2}+\frac{1}{4}}=\sqrt{\frac{\ell\,(\ell+1)-\sigma^2}{\alpha^2}+\frac{1}{4}}.\label{b9}
\end{equation} 

In quantum systems, the requirement of the wave function states that it must be finite and regular everywhere at $r \to 0$ and at $r \to \infty$. Suppose, a possible solution to the Eq. (\ref{b8}) is given by the following form:
\begin{equation}
    \psi(s)=s^{j/2}\,\exp(-s/2)\,F(s),\label{b10}
\end{equation}
where $F(s)$ is an unknown function.

Substituting this solution (\ref{b10}) into the Eq. (\ref{b8}) results the following second-order differential equation form:
\begin{equation}
    s\,F''(s)+(j+1-s)\,F'(s)-\left(j+\frac{1}{2}-\frac{\Lambda}{4\,\Omega}\right)\,F(s)=0.\label{b11}
\end{equation}
This equation (\ref{b11}) is the confluent hypergeometric equation form \cite{MA}, which is a second order linear homogeneous differential equation. Therefore, its regular solution is given by
\begin{equation}
    F(s)={}_1 F_{1} (\mathrm{a}, \mathrm{b}; s),\quad \mathrm{a}=j+\frac{1}{2}-\frac{\Lambda}{4\,\Omega},\quad \mathrm{b}=j+1,\label{b12}
\end{equation}
Due to the asymptotic behavior of this solution, it is necessary that the hypergeometric function be a polynomial function of degree $n$, which means that the parameter $\mathrm{a}$ should be a negative integer. This condition implies that
\begin{equation}
    \left(j+\frac{1}{2}-\frac{\Lambda}{4\,\Omega}\right)=-n\quad\quad (n=0, 1,2,....).\label{b13}
\end{equation}

Simplification of the above relation results the non-relativistic energy eigenvalue given by
\begin{equation}
    E_{n,\ell,\sigma}=2\,\left(n+\sqrt{\frac{\ell\,(\ell+1)-\sigma^2}{\alpha^2}+\frac{1}{4}}+\frac{1}{2}\right)\,\alpha\,\omega.\label{b14}
\end{equation}
The corresponding wave function will be
\begin{eqnarray}
    \psi_{n,\ell,\sigma} (s)&=&s^{\frac{1}{2}\,\sqrt{\frac{\ell\,(\ell+1)-\sigma^2}{\alpha^2}+\frac{1}{4}}}\,\exp(-s/2)\,{}_1 F_{1} \left(-n, 1+\sqrt{\frac{\ell\,(\ell+1)-\sigma^2}{\alpha^2}+\frac{1}{4}}; s\right),\nonumber\\
    R_{n,\ell,\sigma} (s)&=&\Omega^{1/4}\,s^{\frac{1}{2}\,\left(-\frac{1}{2}+\sqrt{\frac{\ell\,(\ell+1)-\sigma^2}{\alpha^2}+\frac{1}{4}}\right)}\,\exp(-s/2)\,{}_1 F_{1} \left(-n, 1+\sqrt{\frac{\ell\,(\ell+1)-\sigma^2}{\alpha^2}+\frac{1}{4}}; s\right)\label{b15}  
\end{eqnarray}

Equation (\ref{b14}) represents the energy spectrum, and Eq. (\ref{b15}) describes the corresponding radial wave function of a harmonic oscillator system in topologically charged space-time and in a Wu-Yang magnetic monopole without EiBI-gravity effects. It is evident that the eigenvalue solution of the harmonic oscillator system is influenced by several factors, including the strength of the WYMM ($\sigma$), the topological parameter $\alpha$, and the oscillator frequency $\omega$. Furthermore, Eq. (\ref{b14}) provides a closed-form expression for the energy spectrum of the harmonic oscillator, which varies with changes in the quantum numbers $(n,\ell)$. 

A special case corresponds to $\sigma=0$, that is without WYMM. Therefore, the energy spectrum (\ref{b14}) and the wave function (\ref{b15}) reduces to 
\begin{eqnarray}
    E_{n,\ell}&=&2\,\left(n+\sqrt{\frac{\ell\,(\ell+1)}{\alpha^2}+\frac{1}{4}}+\frac{1}{2}\right)\,\alpha\,\omega,\nonumber\\
    \psi_{n,\ell} (s)&=&s^{\frac{1}{2}\,\sqrt{\frac{\ell\,(\ell+1)}{\alpha^2}+\frac{1}{4}}}\,\exp(-s/2)\,{}_1 F_{1} \left(-n, 1+\sqrt{\frac{\ell\,(\ell+1)}{\alpha^2}+\frac{1}{4}}; s\right).\label{b16}
\end{eqnarray}
Equation (\ref{b16}) is the eigenvalue solution of a harmonic oscillator in global monopole space-time which is similar to those results obtained in Refs. \cite{ref10, RLLV, CTP}.

\begin{center}
\begin{figure}[ht!]
\begin{centering}
\subfloat[$\omega=1$]{\centering{}\includegraphics[scale=0.4]{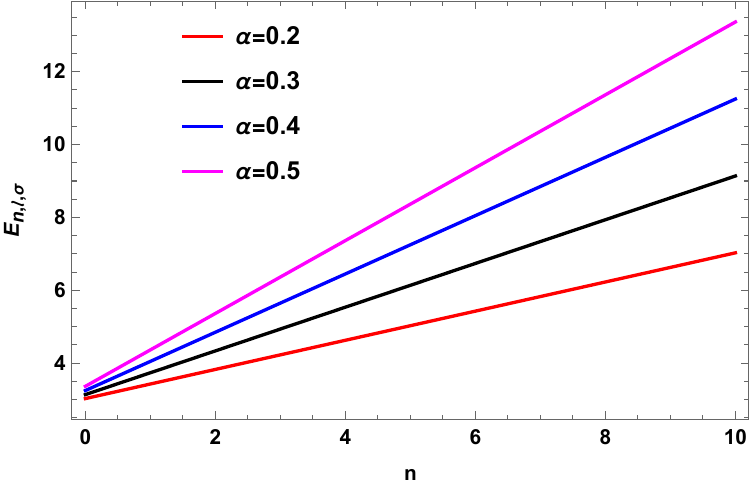}}\quad\quad
\subfloat[$\alpha=0.5$]{\centering{}\includegraphics[scale=0.4]{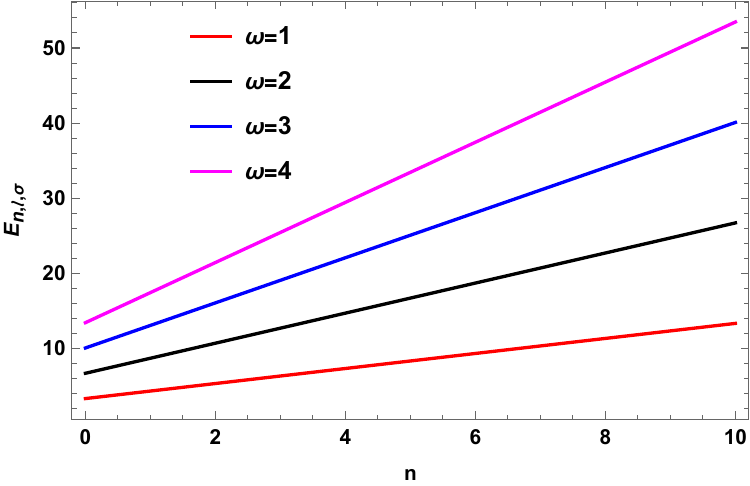}}\quad\quad
\subfloat[$\omega=1$]{\centering{}\includegraphics[scale=0.4]{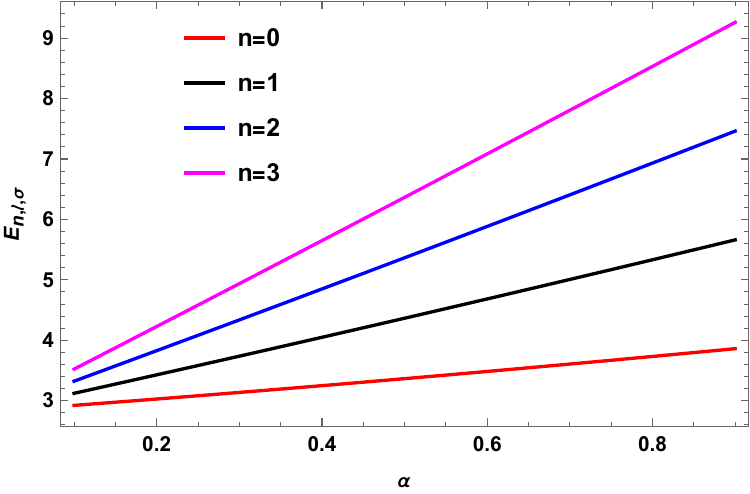}}\quad\quad
\end{centering}
\hfill\\
\begin{centering}
\subfloat[$n=1$]{\centering{}\includegraphics[scale=0.4]{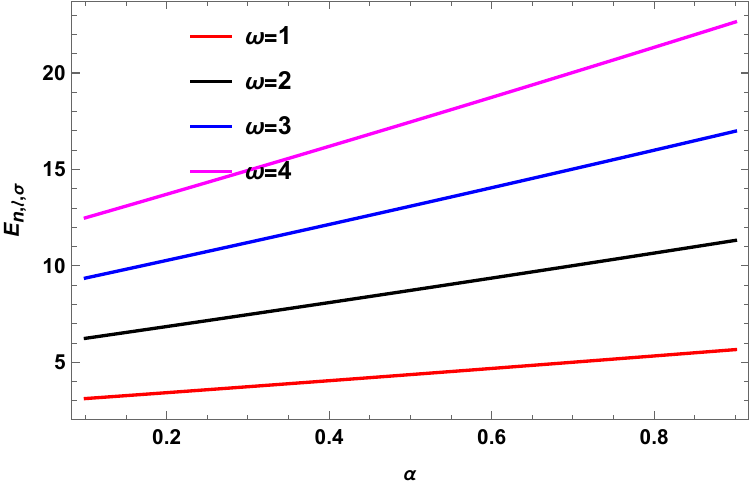}}\quad\quad
\subfloat[$\alpha=0.5$]{\centering{}\includegraphics[scale=0.4]{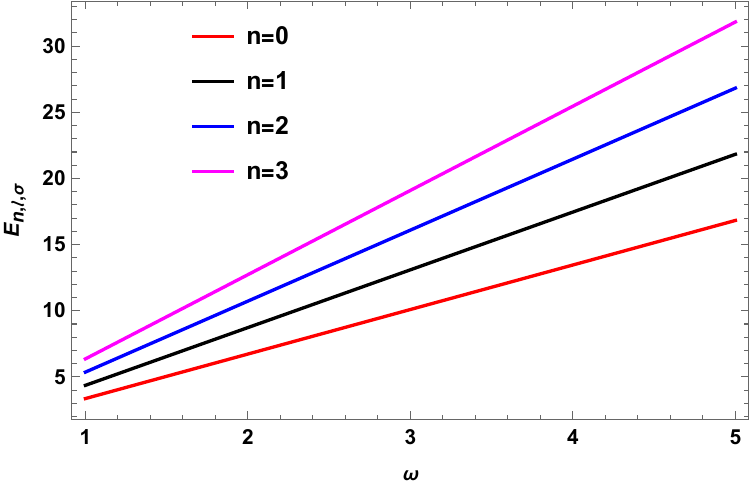}}\quad\quad
\subfloat[$n=1$]{\centering{}\includegraphics[scale=0.4]{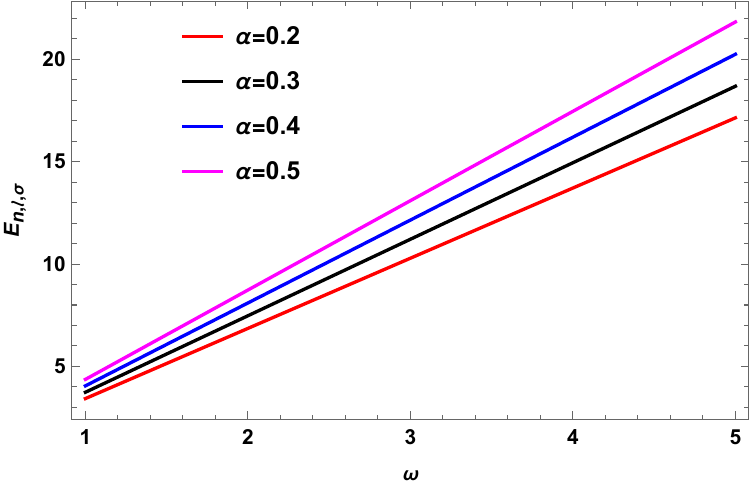}}\quad\quad
\centering{}\caption{The energy level $E_{1,\ell}$ of Eq. (\ref{b14}) for $\ell=1$-state. Here $M=1$, $\kappa=0.5$, $\sigma=0.1$.}\label{fig3}
\end{centering}
\end{figure}
\par\end{center}

\begin{center}
\begin{figure}[ht!]
\subfloat[]{\centering{}\includegraphics[scale=0.4]{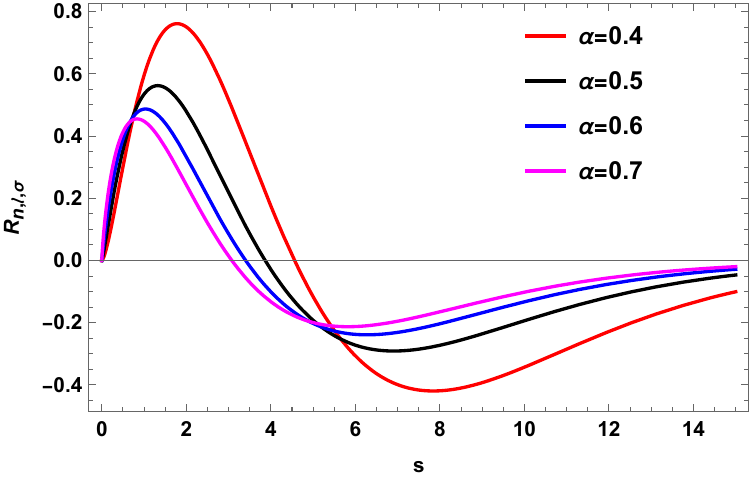}}\quad\quad
\subfloat[]{\centering{}\includegraphics[scale=0.4]{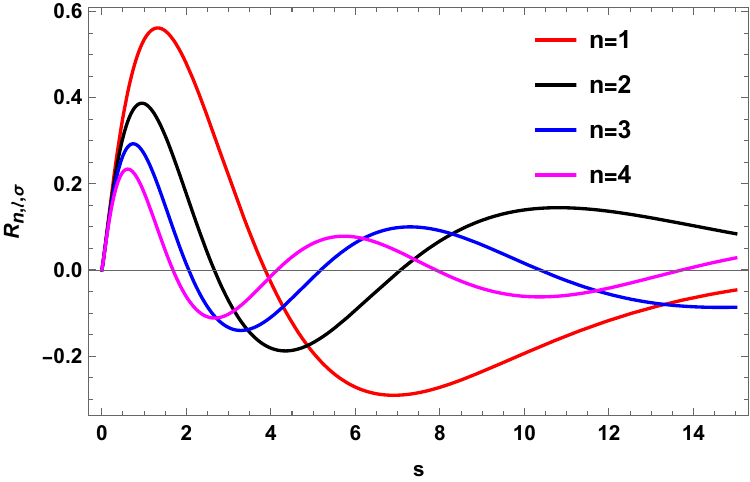}}
\centering{}\caption{The wave function $R_{n,\ell,\sigma}$ of Eq. (\ref{b15}) for $\ell=1$-state. Here $M=1=\omega$, $\sigma=0.1$.}\label{fig-wave-function3}
\end{figure}
\par\end{center}

In Eq. (\ref{b14}), we have seen that the energy expression of the harmonic oscillator is in closed form. To illustrate, we have generated Figure 5 for the $\ell=1$ state, showing the nature of this closed expression of the energy spectrum with respect to $n$, $\alpha$, and $\omega$. In Figures 5(a) and 5(b), we plot $E_{n,\ell}$ versus $n$ for different values of the topology parameter $\alpha$ and the oscillator frequency $\omega$. Figures 5(c) and 5(d) present the graph of $E_{n,\ell}$ versus $\alpha$ for different values of $n$ and $\omega$. Figures 5(e) and 5(f) display the graph of $E_{n,\ell}$ versus $\omega$ for different values of $n$ and $\alpha$. In these graphs, we observe that the energy level increases linearly with $n$, $\alpha$, and $\omega$, with varying slopes. Additionally, this increase shifts upward as the values of the other parameters increase. In Figure 6, we have plotted the radial wave function as given by Eq. (\ref{b15}) for different values of $\alpha$ and $n$, showing its behavior under these variations.

\subsection{Repulsive Inverse Square Potential}

In this part, we want to solve the radial equation (\ref{b6}) for an external potential of the type $V(r)=\frac{\eta}{r^2}$ \cite{ref6} known as inverse square potential, where $\eta>0$ is a constant. So, substituting this potential into the radial Eq. (\ref{b6}), we obtain the following equation:
\begin{eqnarray}
    R^{\prime\prime}(r)+\frac{2}{r}\,R^{\prime} (r)+\Bigg[\frac{2\,M\,E}{\alpha^2}-\Omega^2\,r^2-\frac{\lambda^{\prime}+2\,M\,\eta}{\alpha^2\,r^2} \Bigg]\,R(r)=0.\label{d1}
\end{eqnarray}

Transforming first $R(r)=\frac{\psi(r)}{\sqrt{r}}$, and then to a new variable $s=\Omega\,r^2$ into the Eq. (\ref{d1}), we obtain the following differential equation form: 
\begin{equation}
    \psi''(s)+\frac{1}{s}\,\psi'(s)+\Bigg[-\frac{\tau^2/4}{s^2}-\frac{1}{4}+\frac{\Lambda}{4\,\Omega\,s}  \Bigg]\,\psi(s)=0,\label{d2}
\end{equation}
where $\Lambda$ is defined earlier and
\begin{equation}
    \tau=\sqrt{\frac{\ell\,(\ell+1)-\sigma^2+2\,M\,\eta}{\alpha^2}+\frac{1}{4}}.\label{d3}
\end{equation}

Analogue to the previous analysis, suppose a possible solution to the Eq. (\ref{d2}) is of the following form
\begin{equation}
    \psi(s)=s^{\tau/2}\,\exp(-s/2)\,F(s),\label{d4}
\end{equation}
where $F(s)$ is an unknown function.

Substituting this solution (\ref{d4}) into the Eq. (\ref{d2}) results the following second-order differential equation:
\begin{equation}
    s\,F''(s)+(\tau+1-s)\,F'(s)-\left(\tau+\frac{1}{2}-\frac{\Lambda}{4\,\Omega}\right)\,F(s)=0.\label{d5}
\end{equation}
The above differential equation is the confluent hypergeometric equation form \cite{MA}. Therefore, its regular solution is given by
\begin{equation}
    F(s)={}_1 F_{1} (\mathrm{p}, \mathrm{q}; s),\quad\quad \mathrm{p}=\tau+\frac{1}{2}-\frac{\Lambda}{4\,\Omega},\quad\quad \mathrm{q}=\tau+1,\label{d6}
\end{equation}
Due to the asymptotic behavior of this solution, it is necessary that the hypergeometric function be a polynomial function of degree $n$, which means that the parameter $\mathrm{p}$ should be a negative integer. This condition implies that
\begin{equation}
    \left(\tau+\frac{1}{2}-\frac{\Lambda}{4\,\Omega}\right)=-n\quad\quad (n=0, 1,2,....).\label{d7}
\end{equation}

Simplification of the above relation results the non-relativistic energy eigenvalue of the harmonic oscillator system given by
\begin{equation}
    E_{n,\ell,\sigma}=2\,\left(n+\sqrt{\frac{\ell\,(\ell+1)-\sigma^2+2\,M\,\eta}{\alpha^2}+\frac{1}{4}}+\frac{1}{2}\right)\,\alpha\,\omega.\label{d8}
\end{equation}
The corresponding wave function will be
\begin{eqnarray}
    \psi_{n,\ell,\sigma} (s)&=&s^{\frac{1}{2}\,\sqrt{\frac{\ell\,(\ell+1)-\sigma^2+2\,M\,\eta}{\alpha^2}}}\,\exp(-s/2)\,{}_1 F_{1} \left(-n, 1+\sqrt{\frac{\ell\,(\ell+1)-\sigma^2+2\,M\,\eta}{\alpha^2}}; s\right),\nonumber\\
    R_{n,\ell,\sigma} (s)&=&\Omega^{1/4}\,s^{\frac{1}{2}\,\left(-\frac{1}{2}+\sqrt{\frac{\ell\,(\ell+1)-\sigma^2+2\,M\,\eta}{\alpha^2}+\frac{1}{4}}\right)}\,\exp(-s/2)\,{}_1 F_{1} \left(-n, 1+\sqrt{\frac{\ell\,(\ell+1)-\sigma^2+2\,M\,\eta}{\alpha^2}+\frac{1}{4}}; s\right).\label{d9}
\end{eqnarray}

Equation (\ref{d8}) is the non-relativistic energy spectrum, while (\ref{d9}) describes the corresponding radial wave function of a harmonic oscillator in the context of global monopole (GM) space-time and in a WYMM without EiBI-gravity effects, under the influence of an external inverse square potential. The eigenvalue solution of the harmonic oscillator system is influenced by several factors, including the strength of the WYMM ($\sigma$), the topological parameter ($\alpha$), the oscillator frequency ($\omega$), and the potential parameter ($\eta$). Additionally,  Eq. (\ref{d8}) provides a closed-form expression for the energy spectrum of the harmonic oscillator, which varies with the quantum numbers $\{n,\ell\}$.

A special case corresponds to $\sigma=0$, which implies the absence of WYMM. It means the harmonic oscillator system under the influence of an inverse square potential in GM space-time, without EiBI-gravity and WYMM effects. In other words, it describes the non-relativistic quantum particles interacting with a pseudoharmonic-type potential in a GM space-time background, devoid of additional effects such as EiBI-gravity and WYMM. Consequently, the non-relativistic energy spectrum given by Eq. (\ref{d8}) and the wave function described by Eq. (\ref{d9}) reduce to the following form:
\begin{eqnarray}
    E_{n,\ell}&=&2\,\left(n+\sqrt{\frac{\ell\,(\ell+1)+2\,M\,\eta}{\alpha^2}+\frac{1}{4}}+\frac{1}{2}\right)\,\alpha\,\omega\,,\nonumber\\
    \psi_{n,\ell} (s)&=&s^{\frac{1}{2}\,\sqrt{\frac{\ell\,(\ell+1)+2\,M\,\eta}{\alpha^2}}}\,\exp(-s/2)\,{}_1 F_{1} \left(-n, 1+\sqrt{\frac{\ell\,(\ell+1)+2\,M\,\eta}{\alpha^2}}; s\right)\,.\label{d10}
\end{eqnarray}

Equation (\ref{d10}) represents the non-relativistic energy spectrum of a harmonic oscillator influenced by an inverse square potential in global monopole (GM) space-time, excluding the effects of EiBI-gravity and the Weinberg-Yang-Mills model (WYMM). This result is consistent with those obtained in Ref. \cite{CTP}. Consequently, it is clear that the presence of a WYMM in the harmonic oscillator system significantly affects both the energy spectrum (\ref{d8}) and the wave function (\ref{d9}), alongside other parameters considered in this study.  

\begin{center}
\begin{figure}[ht!]
\subfloat[$\omega=1=\eta$]{\centering{}\includegraphics[scale=0.4]{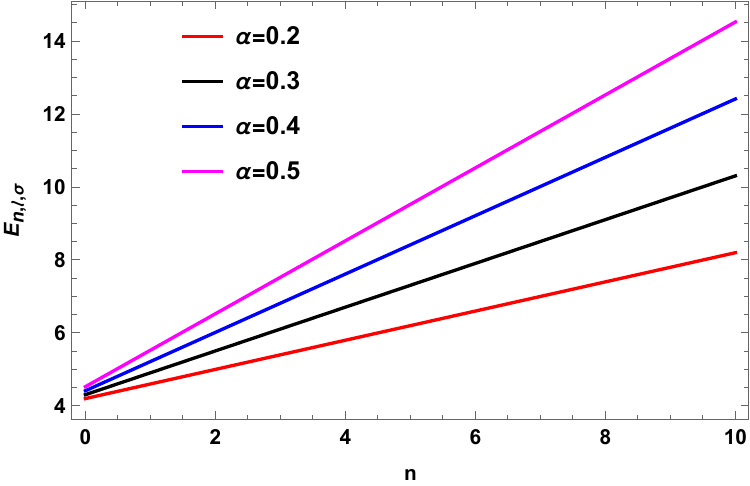}}\quad\quad
\subfloat[$\eta=1$, $\alpha=0.5$]{\centering{}\includegraphics[scale=0.4]{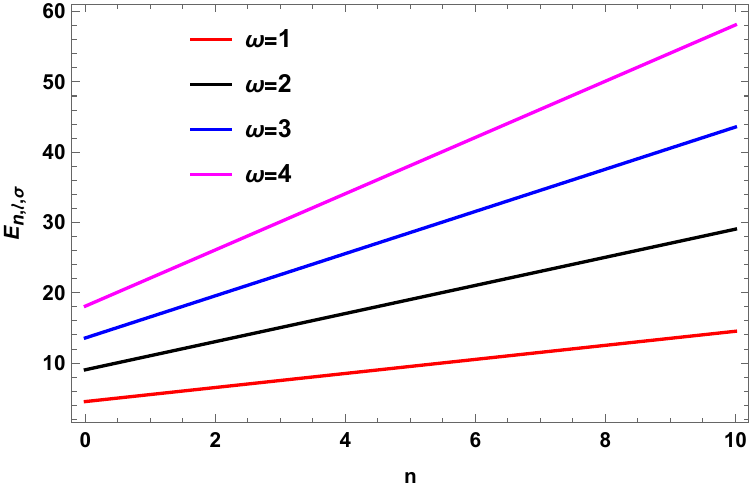}}\quad\quad
\subfloat[$\omega=1$, $\alpha=0.5$]{\centering{}\includegraphics[scale=0.4]{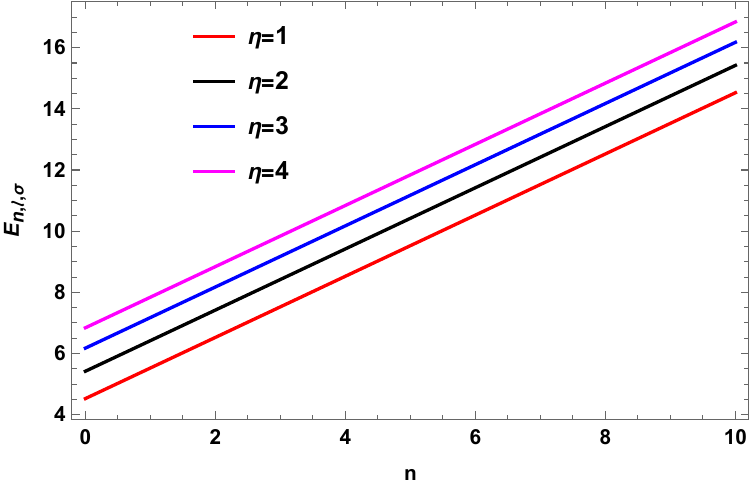}}
\hfill\\
\subfloat[$\omega=1=\eta$]{\centering{}\includegraphics[scale=0.4]{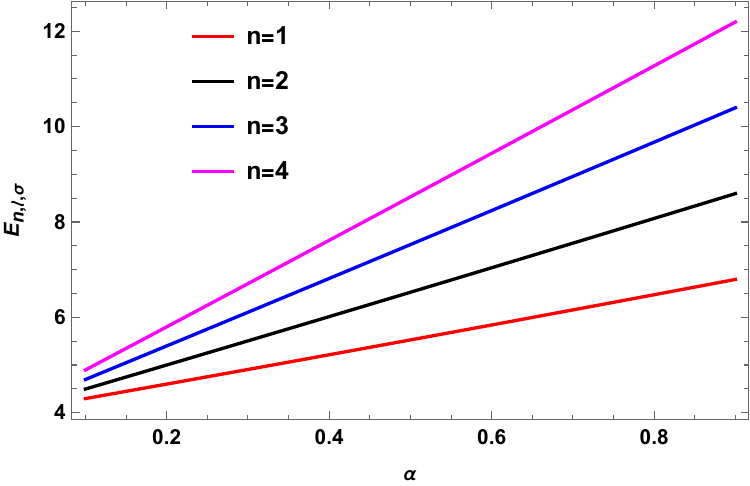}}\quad\quad
\subfloat[$n=1$, $\eta=0.5$]{\centering{}\includegraphics[scale=0.4]{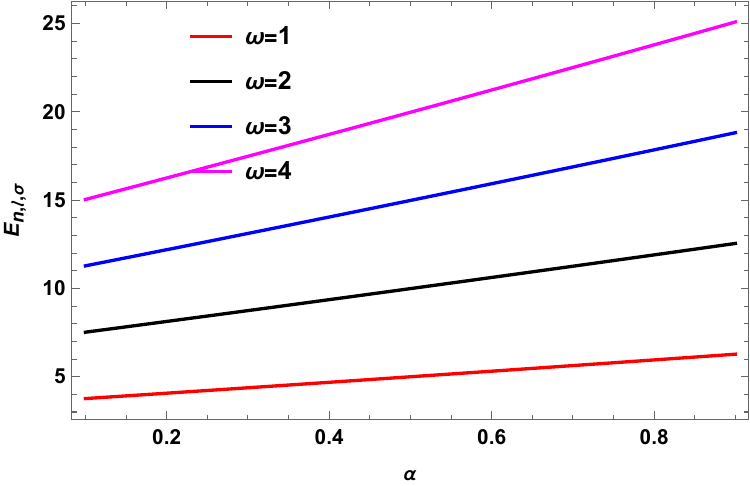}}\quad\quad
\subfloat[$n=1=\omega$]{\centering{}\includegraphics[scale=0.4]{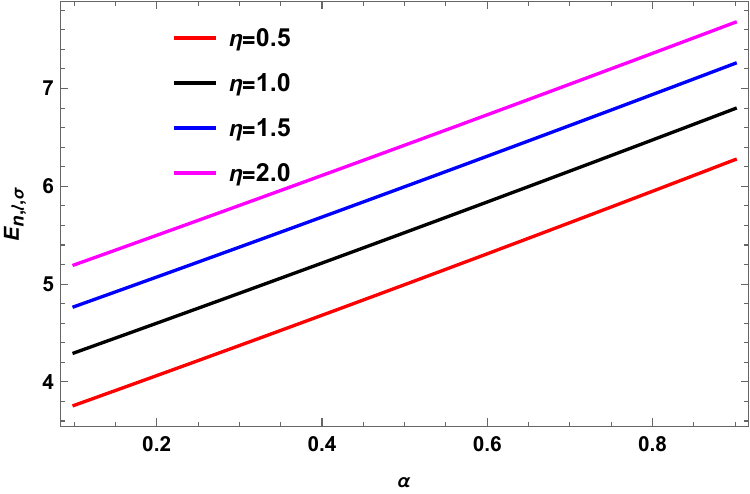}}
\hfill\\
\subfloat[$\alpha=0.5$, $\eta=1$]{\centering{}\includegraphics[scale=0.4]{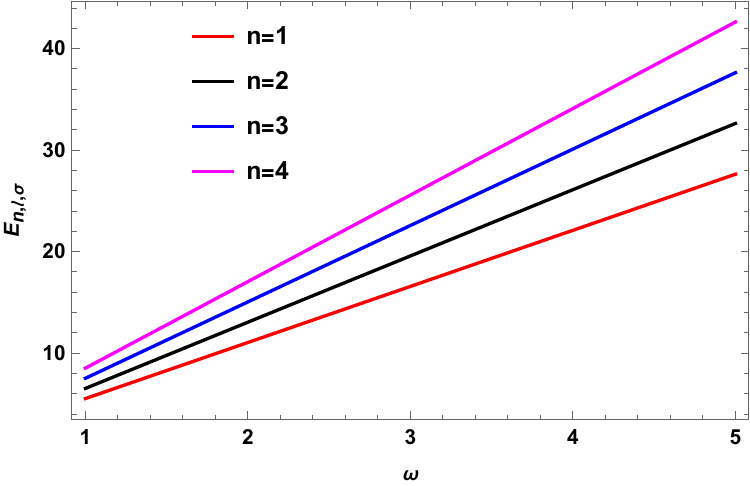}}\quad\quad
\subfloat[$n=1$, $\eta=1$]{\centering{}\includegraphics[scale=0.4]{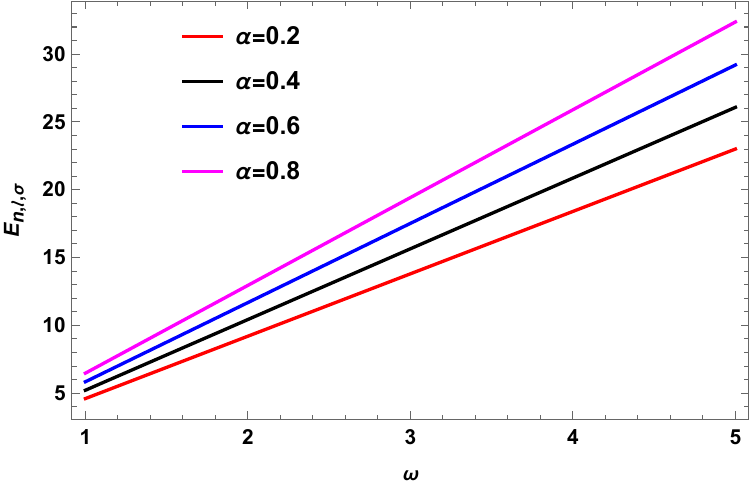}}\quad\quad
\subfloat[$n=1$, $\alpha=0.5$]{\centering{}\includegraphics[scale=0.4]{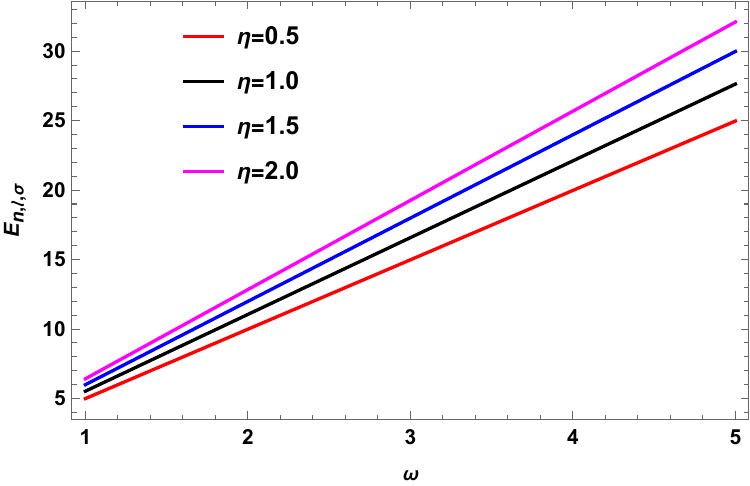}}
\centering{}\caption{The energy level $E_{n,\ell,\sigma}$ of Eq. (\ref{d8}) for $\ell=1$-state. Here $M=1$, $\kappa=0.5$, $\sigma=0.1$.}\label{fig4}
\end{figure}
\par\end{center}

\begin{center}
\begin{figure}[ht!]
\subfloat[$n=1$, $\eta=0.5$]{\centering{}\includegraphics[scale=0.32]{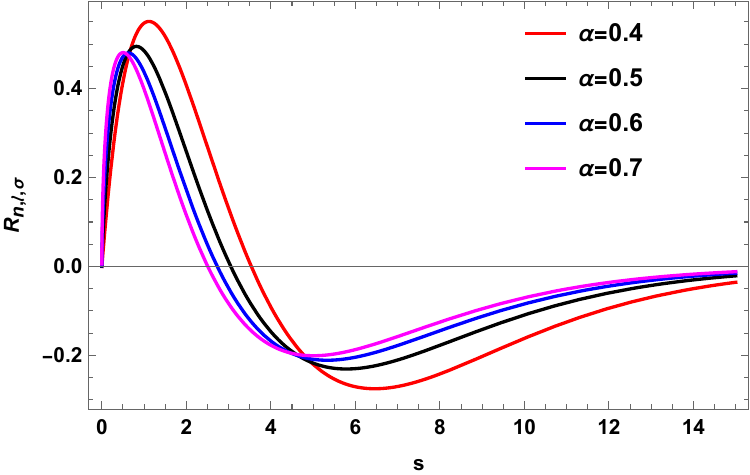}}\quad
\subfloat[$\alpha=0.5=\eta$]{\centering{}\includegraphics[scale=0.32]{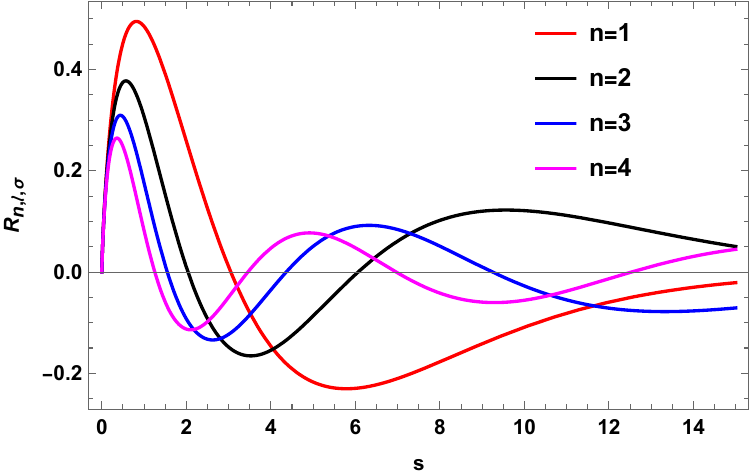}}\quad
\subfloat[$n=1$, $\alpha=0.5$]{\centering{}\includegraphics[scale=0.32]{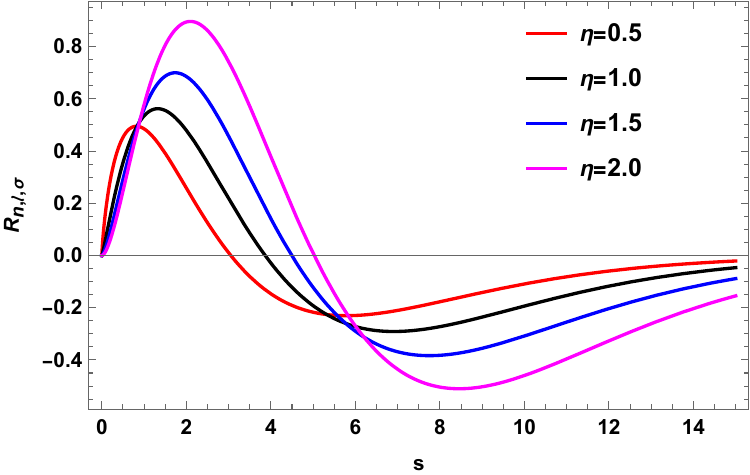}}\quad
\subfloat[$n=1$]{\centering{}\includegraphics[scale=0.32]{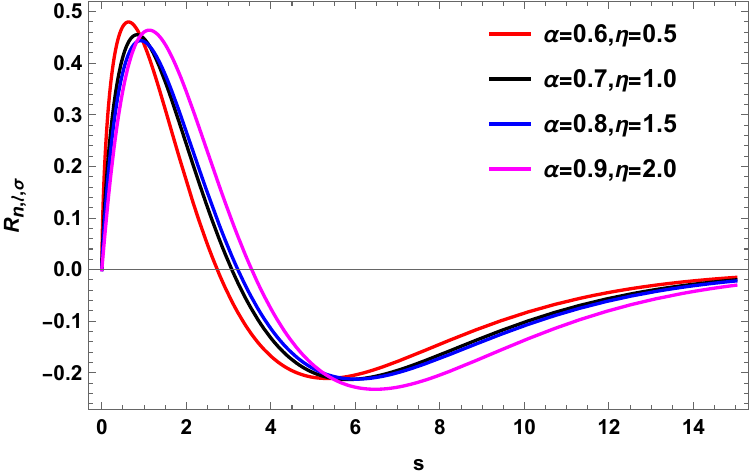}}
\centering{}\caption{The wave function $R_{n,\ell,\sigma}$ of Eq. (\ref{d9}) for $\ell=0$-state. $M=1=\omega$, $\sigma=0.1$.}\label{fig-wave-function4}
\hfill\\
\subfloat[$n=1$, $\eta=0.5$]{\centering{}\includegraphics[scale=0.32]{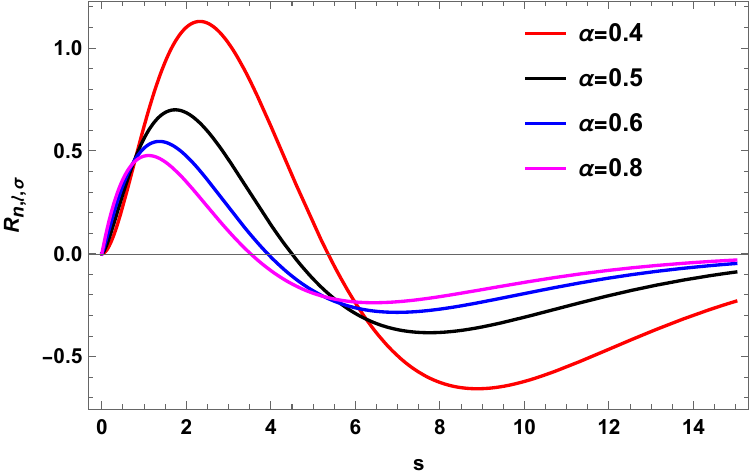}}\quad
\subfloat[$\alpha=0.5=\eta$]{\centering{}\includegraphics[scale=0.32]{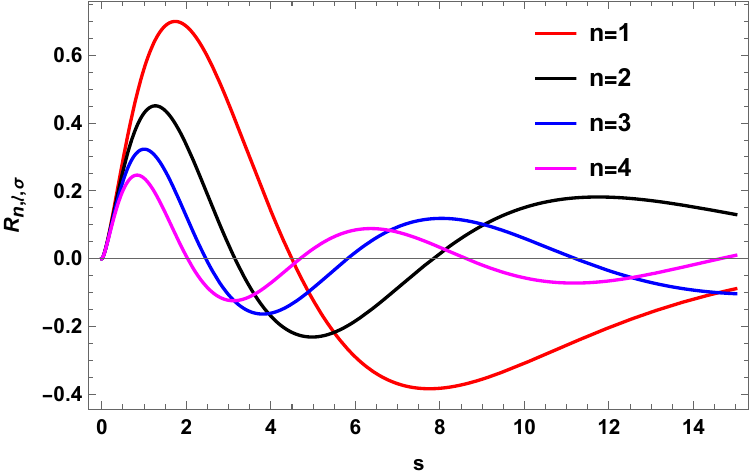}}\quad
\subfloat[$n=1$, $\alpha=0.5$]{\centering{}\includegraphics[scale=0.32]{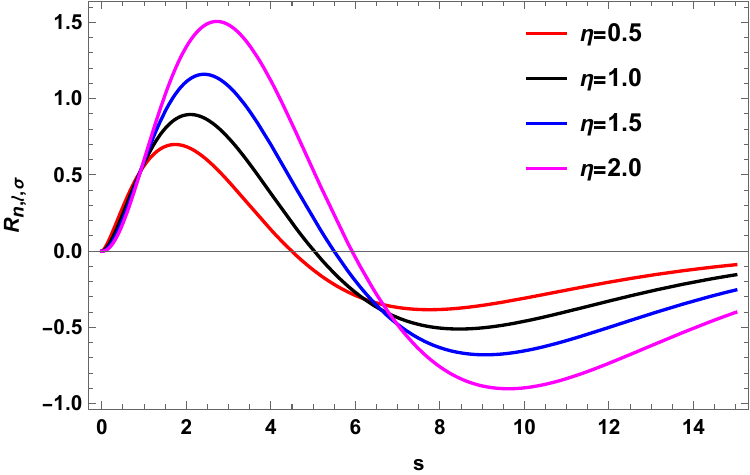}}\quad
\subfloat[$n=1$]{\centering{}\includegraphics[scale=0.32]{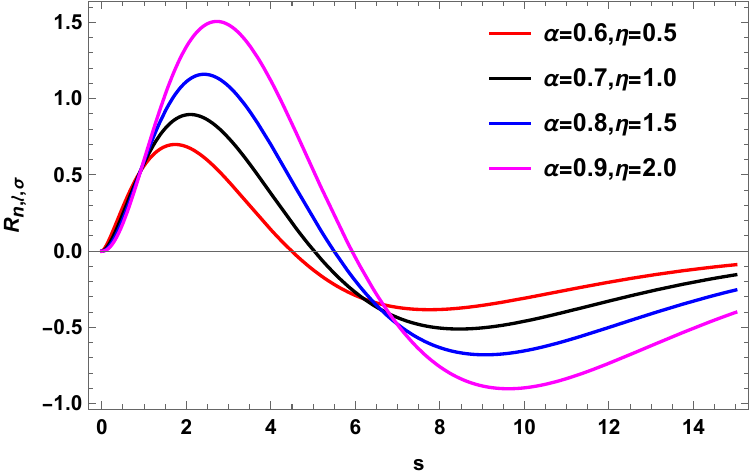}}
\centering{}\caption{The wave function $R_{n,\ell,\sigma}$ of Eq. (\ref{d9}) for $\ell=1$-state. $M=1=\omega$, $\sigma=0.1$.}\label{fig-wave-function5}
\end{figure}
\par\end{center}

In Eq. (\ref{d8}), the energy spectrum is expressed in a closed or compact form and varies with the quantum numbers $(n, \ell)$. For $\ell=1$-state solution, we have generated Figures 7(a) through 7(i) to illustrate this energy spectrum as functions of $n$, $\alpha$, and $\omega$. We observe that the energy spectrum for a particular state linearly increases with different slopes when plotted against these parameters. Additionally, as the values of the other parameters increase, this linear increase shifts the energy level upward. While we have focused on the $\ell=1$ state, one could also explore the $\ell=0$ state, which we have left for future investigation.

In Figures 8 and 9, we have plotted the radial wave function as given by Eq. (\ref{d9}), varying the values of different parameters involved for both the $\ell=0$ and $\ell=1$-states. These plots demonstrate the behavior of the radial wave function under various conditions.

\section{Harmonic oscillator system in topologically charged EiBI-gravity space-time and in a WYMM}

In this section, we now study the harmonic oscillator system in topologically charged space-time within the context of EiBI-gravity and in a WYMM. Therefore, the radial equation describing harmonic oscillator system in absence of external potential $V(\rho)=0$ in this environment is given by
\begin{eqnarray}
    \Bigg[\frac{d^2}{d\rho^2}+\frac{2\,\rho}{(\rho^2-\kappa)}\,\frac{d}{d\rho}+\frac{2\,M\,E}{\alpha^2}-\frac{M^2\,\omega^2}{\alpha^2}\,\rho^2-\frac{\lambda^{\prime}/\alpha^2}{(\rho^2-\kappa)}\Bigg]\,R(\rho)=0,\label{ee1}
\end{eqnarray}

Transforming to a new function defined in Eq. (\ref{a9}) into the Eq. (\ref{ee1}) results
\begin{eqnarray}
    \psi''(\rho)+\Bigg[\frac{2\,\rho}{\rho^2-\kappa}-2\,\Omega\,\rho\Bigg]\,\psi'(\rho)+\Bigg[\Delta-\frac{2\,\Omega\,\rho^2}{\rho^2-\kappa}-\frac{\iota^2}{\rho^2-\kappa}\Bigg]\,\psi(\rho)=0,\label{ee2}
\end{eqnarray}
where $\Delta, \Omega$ are given in Eq. (\ref{a11}) and 
\begin{equation}
    \iota=\sqrt{\frac{\ell\,(\ell+1)-\sigma^2}{\alpha^2}}.\label{ee3}
\end{equation}

Finally transforming to a new variable via $\rho^2=s\,\kappa$ into the equation (\ref{ee2}) results the following differential equation form
\begin{eqnarray}
    \psi''(s)+\left[-\Omega\,\kappa+\frac{1/2}{s}+\frac{1}{s-1}  \right]\,\psi'(s)+\Bigg[\frac{\left(\Delta\,\kappa+\iota^2\right)/4}{s}+\frac{-\left(2\,\Omega\,\kappa+\iota^2\right)/4}{s-1} \Bigg]\,\psi(s)=0.\label{ee4}
\end{eqnarray}

Comparing Eq. (\ref{ee4}) with Eq. (A1) in appendix, we obtain different parameters
\begin{equation}
    \Xi=-\Omega\,\kappa,\quad\quad \beta=-1/2,\quad\quad \gamma=0,\quad\quad \tilde{\mu}=\left(\Delta\,\kappa+\iota^2\right)/4,\quad\quad \tilde{\nu}=-\left(2\,\Omega\,\kappa+\iota^2\right)/4.\label{ee5}
\end{equation}
Thus, $\psi (s)$ is the confluent Heun function given by
\begin{equation}
    \psi(s)=H_{c}\left(-\Omega\,\kappa, -\frac{1}{2}, 0, \frac{M\,E\,\kappa}{2\,\alpha^2}, \frac{1}{4}-\frac{1}{4\,\alpha^2}\left(2\,M\,E\,\kappa+\ell\,(\ell+1)-\sigma^2\right) ;s\right).\label{ee6}
\end{equation}

In order to solve q. (\ref{ee4}), we consider a power series expansion given by $H_{c}(s)=\sum^{\infty}_{i=0}\,d_{i}\,s^{i}$ \cite{GBA}. Substituting this power series into the Eq. (\ref{ee4}), we obtain following recurrence relation
\begin{eqnarray}
    d_{i+2}=\frac{1}{(i+2)(2\,i+3)}\Big[\Big\{(i+1)(2\,i+3+2\,\Omega\,\kappa)-2\,\tilde{\mu}\Big\}\,d_{i+1}+2\,(-\Omega\,\kappa\,i+\tilde{\mu}+\tilde{\nu})\,d_{i}\Big] \label{ee7}
\end{eqnarray}
with a few coefficients given by
\begin{eqnarray}
    d_1&=&-2\,\tilde{\mu}\,d_0,\nonumber\\
    d_2&=&\frac{1}{6}\Big[\Big\{(3+2\,\Omega\,\kappa)-2\,\tilde{\mu}\Big\}\,d_{1}+2\,(\tilde{\mu}+\tilde{\nu})\,d_{0}\Big].\label{ee8}
\end{eqnarray}

Analogue to the previous analysis done in section 2, we now truncate the power series to a finite degree polynomial of order $n$. This truncation is achieved by ensuring the coefficient $d_{n+1}=0$ in the recurrence relation (\ref{ee7}), where $i=(n-1)$. Consequently, the recurrence relation Eq. (\ref{ee7}) under this condition can be rewritten as follows:
\begin{eqnarray}
    d_n=-\frac{2\,\Big[-\Omega\,\kappa\,(n-1)+\tilde{\mu}+\tilde{\nu}\Big]}{\Big[n\,(2\,n+1+2\,\Omega\,\kappa)-2\,\tilde{\mu}\Big]}\,d_{n-1},\label{ee9}
\end{eqnarray}
where $n=1,2,3,...$ represent the radial modes of the quantum system. 

As done in the earlier section, let us consider the radial mode $n=1$, which represents the lowest energy state of the quantum system. Therefore, substituting $n=1$ into Eq. (\ref{ee9}), we obtain the following expression
\begin{eqnarray}
    d_1=-\frac{2\,(\tilde{\mu}+\tilde{\nu})}{\Big[3+2\,\Omega\,\kappa-2\,\tilde{\mu}\Big]}\,\,d_{0}.\label{ee10}
\end{eqnarray}

Thus, by combining Eqs. (\ref{ee8}) and (\ref{ee10}) gives us the following expression of the energy eigenvalues given by
\begin{eqnarray}
    E_{1,\ell,\sigma}&=&\frac{1}{2\,M}\Bigg[3\,M\,\omega\,\alpha-\frac{\ell\,(\ell+1)}{\kappa}+\frac{\sigma^2}{\kappa}+\frac{2\,\alpha^2}{\kappa}\nonumber\\
    &&\pm\,\frac{\alpha^2}{\kappa}\,\sqrt{\frac{2\,\ell\,(\ell+1)-2\,\sigma^2}{\alpha^2}+4+\frac{4\,M\,\omega\,\kappa}{\alpha}\,\left(\frac{M\,\omega\,\kappa}{\alpha}+4\right)}\Bigg].\label{ee11}
\end{eqnarray}
Equation (\ref{ee11}) represents the allowed energy values for the lowest energy state of harmonic oscillator in topologically charged EiBI-gravity space-time and in a WYMM.

The eigenfunction corresponds to the ground state of the energy spectrum Eq. (\ref{ee11}) is described by the first term of the polynomial of the confluent Heun equation Eq. (\ref{ee6}), defined as $\psi_{1,\ell,\sigma}(s)=(d_0+d_1\,s)$. Therefore, the ground state wave function is given by 
\begin{eqnarray}
    R_{1,\ell,\sigma} (\rho)&=&\exp\left(-\frac{M\,\omega}{2\,\alpha}\,\rho^2\right)\,\Bigg[1-\rho^2\,\Bigg\{\frac{1}{\kappa}+\frac{M\,\omega}{\alpha}\pm\frac{1}{2\,\kappa}\times\nonumber\\
    &&\sqrt{\frac{2\,\ell\,(\ell+1)-2\,\sigma^2}{\alpha^2}+4+\frac{4\,M\,\omega\,\kappa}{\alpha}\,\left(\frac{M\,\omega\,\kappa}{\alpha}+4\right)}\Bigg\}\Bigg]\,d_0.\label{ee12}
\end{eqnarray}

Now, we discuss a special case corresponds to zero oscillator frequency, $\omega=0$. So, by making $\omega=0$ in the energy spectrum Eq. (\ref{ee11}), we obtain
\begin{equation}
    E_{1,\ell,\sigma}=\frac{1}{2\,M}\Bigg[-\frac{\ell\,(\ell+1)}{\kappa}+\frac{\sigma^2}{\kappa}+\frac{2\,\alpha^2}{\kappa}\pm\,\frac{\alpha^2}{\kappa}\,\sqrt{\frac{2\,\ell\,(\ell+1)-2\,\sigma^2}{\alpha^2}+4}\Bigg].\label{ee13}
\end{equation}
That is, the allowed energy values for the lowest energy state of a non-relativistic quantum particle in topologically charged EiBI-gravity space-time and in a WYMM. This means that, even without harmonic interaction, the non-relativistic quantum particle continues to have discrete energy, characterizing the confinement, which comes from the gravitational effects. The corresponding eigenfunction in this special case from Eq. (\ref{ee12}) reduces as follows:
\begin{equation}
    R_{1,\ell,\sigma} (\rho)=\Bigg[1-\rho^2\,\left\{\frac{1}{\kappa}\pm\frac{1}{2\,\kappa}\,\sqrt{\frac{2\,\ell\,(\ell+1)-2\,\sigma^2}{\alpha^2}+4}\right\}\Bigg]\,d_0.\label{ee14}
\end{equation}

\begin{center}
\begin{figure}[ht!]
\subfloat[]{\centering{}\includegraphics[scale=0.32]{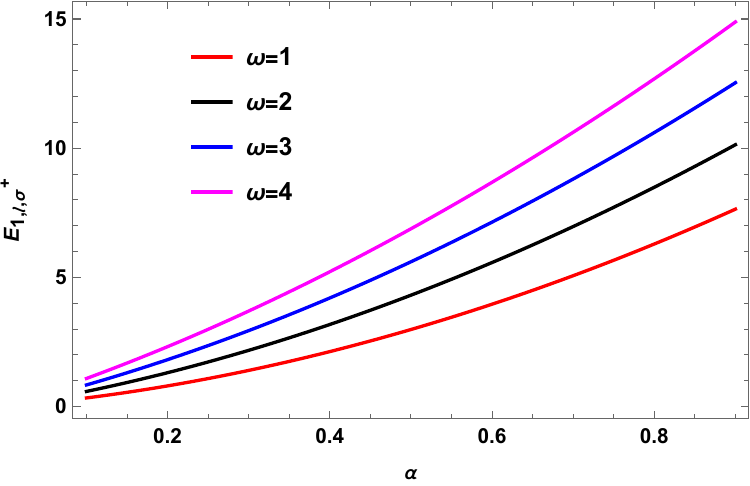}}\quad
\subfloat[]{\centering{}\includegraphics[scale=0.32]{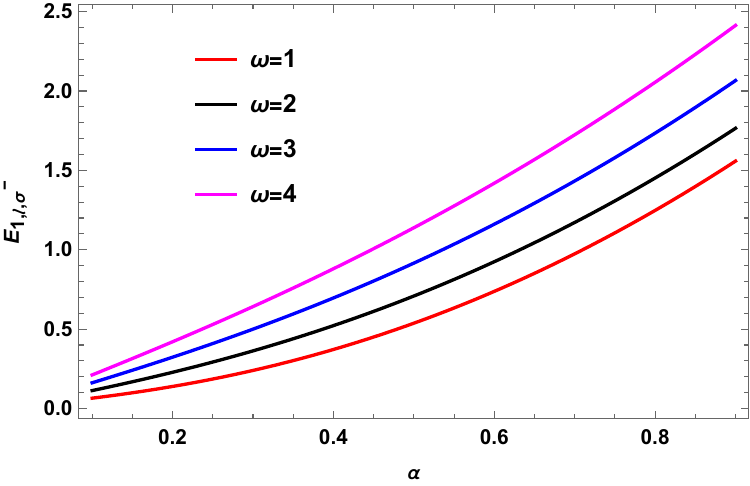}}\quad
\subfloat[]{\centering{}\includegraphics[scale=0.32]{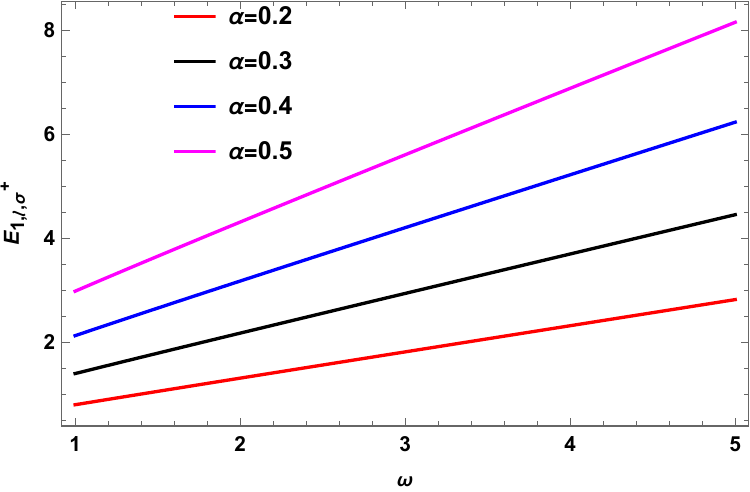}}\quad
\subfloat[]{\centering{}\includegraphics[scale=0.32]{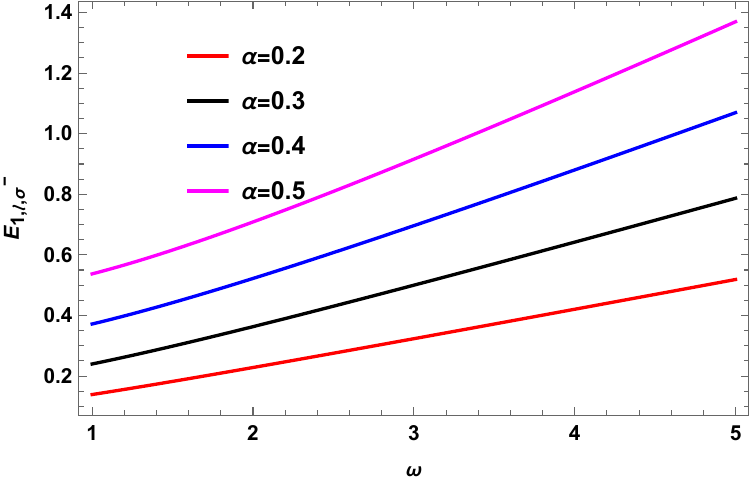}}
\centering{}\caption{The energy level $E_{1,\ell}$ of Eq. (\ref{ee11}) for $\ell=0$-state. Here, $E^{+}$ indicates positive sign within the energy expression and $E^{-}$ that for negative sign. Here $M=1$, $\kappa=0.5$, $\sigma=0.1$.}\label{fig7}
\end{figure}
\par\end{center}

\begin{center}
\begin{figure}[ht!]
\subfloat[]{\centering{}\includegraphics[scale=0.32]{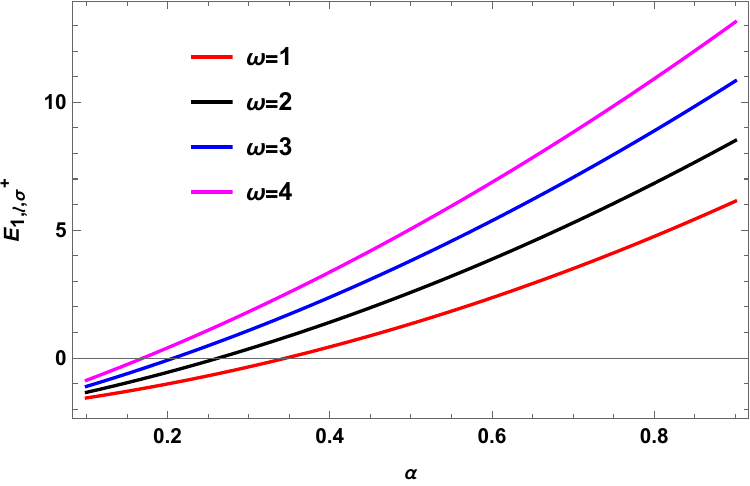}}\quad
\subfloat[]{\centering{}\includegraphics[scale=0.32]{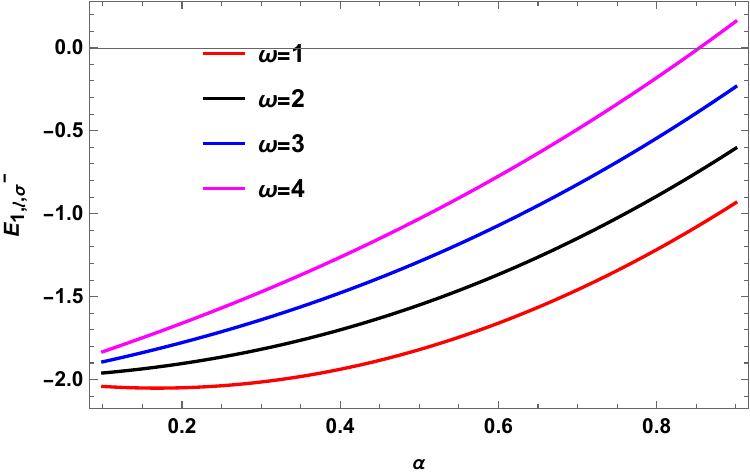}}\quad
\subfloat[]{\centering{}\includegraphics[scale=0.32]{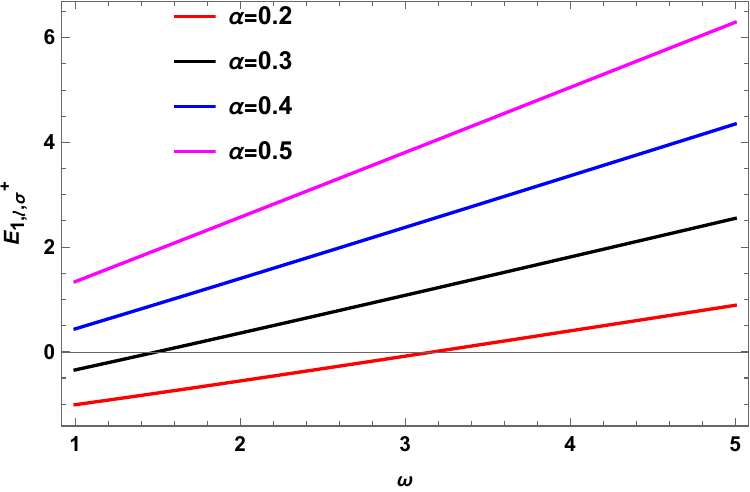}}\quad
\subfloat[]{\centering{}\includegraphics[scale=0.32]{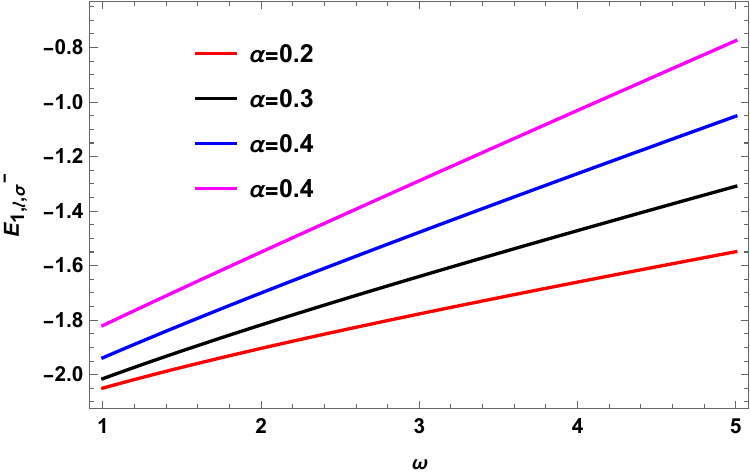}}
\centering{}\caption{The energy level $E_{1,\ell}$ of Eq. (\ref{ee11}) for $\ell=1$-state. Here, $E^{+}$ indicates positive sign within the energy expression and $E^{-}$ that for negative sign. Here $M=1$, $\kappa=0.5$, $\sigma=0.1$.}\label{fig8}
\end{figure}
\par\end{center}

\begin{center}
\begin{figure}[ht!]
\subfloat[$\omega=1$]{\centering{}\includegraphics[scale=0.32]{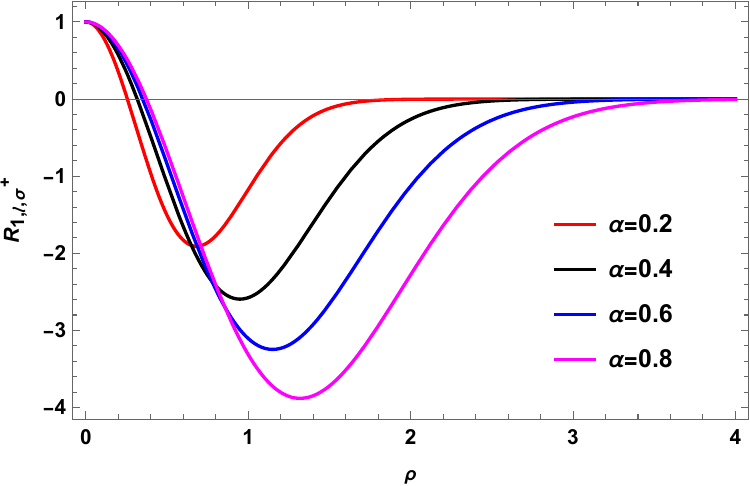}}\quad
\subfloat[$\omega=1$]{\centering{}\includegraphics[scale=0.32]{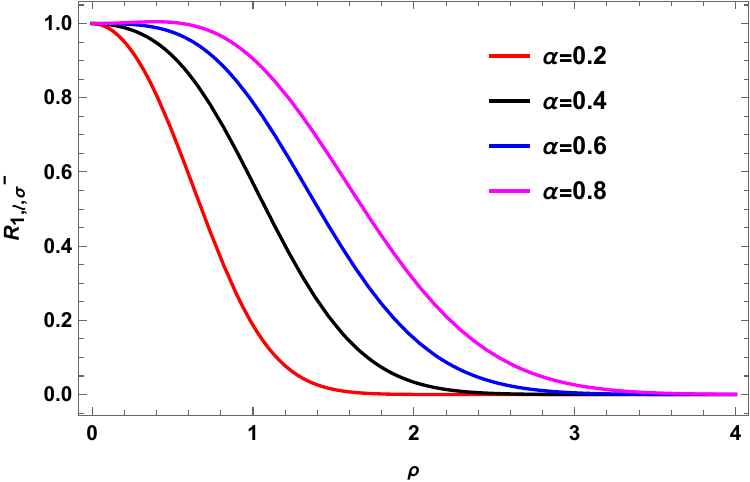}}\quad
\subfloat[$\alpha=0.5$]{\centering{}\includegraphics[scale=0.32]{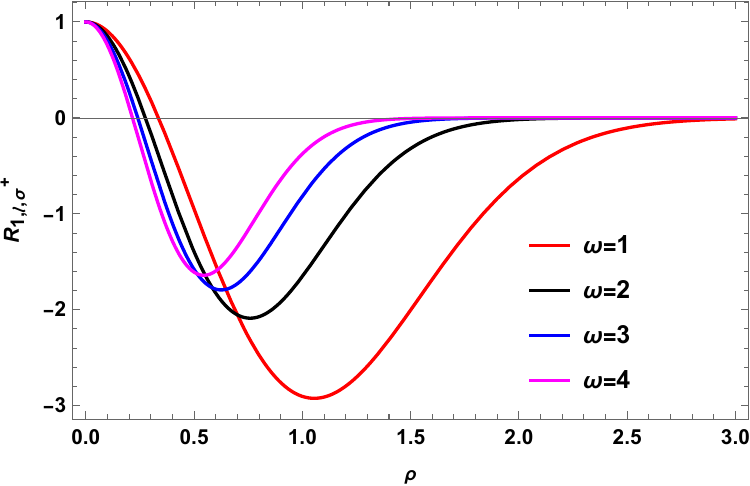}}\quad
\subfloat[$\alpha=0.5$]{\centering{}\includegraphics[scale=0.32]{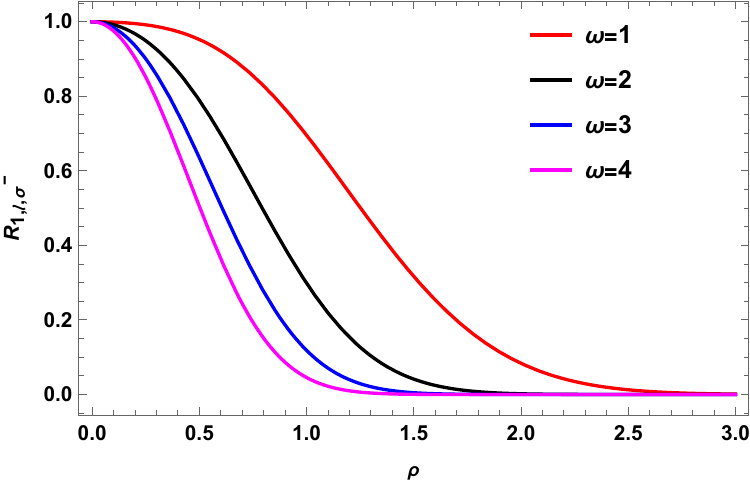}}
\centering{}\caption{The wave function $R_{1,\ell,\sigma}$ of Eq. (\ref{ee12}) for $\ell=0$-state. Here, $R^{+}$ indicates positive sign within the expression and $R^{-}$ that for negative sign.  Here $M=1$, $\kappa=0.5$, $\sigma=0.1$.}\label{fig-wave-function6}
\hfill\\
\subfloat[$\omega=1$]{\centering{}\includegraphics[scale=0.32]{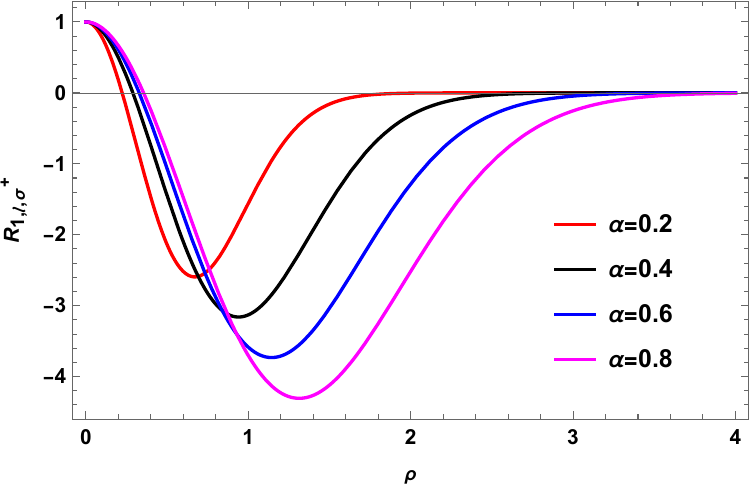}}\quad
\subfloat[$\omega=1$]{\centering{}\includegraphics[scale=0.32]{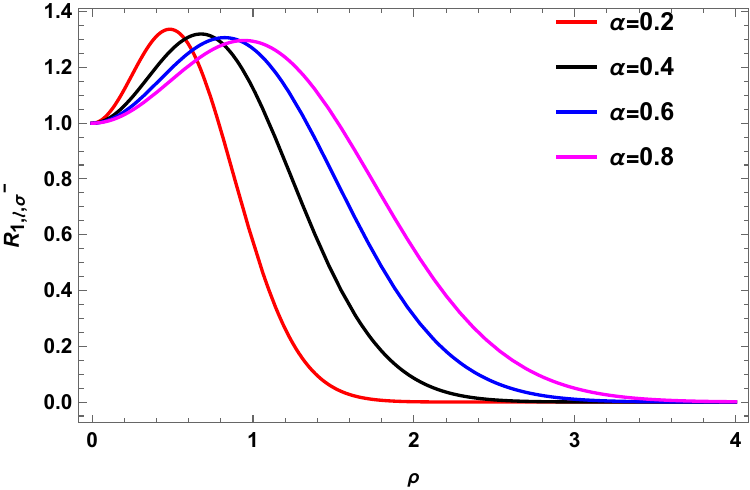}}\quad
\subfloat[$\alpha=0.5$]{\centering{}\includegraphics[scale=0.32]{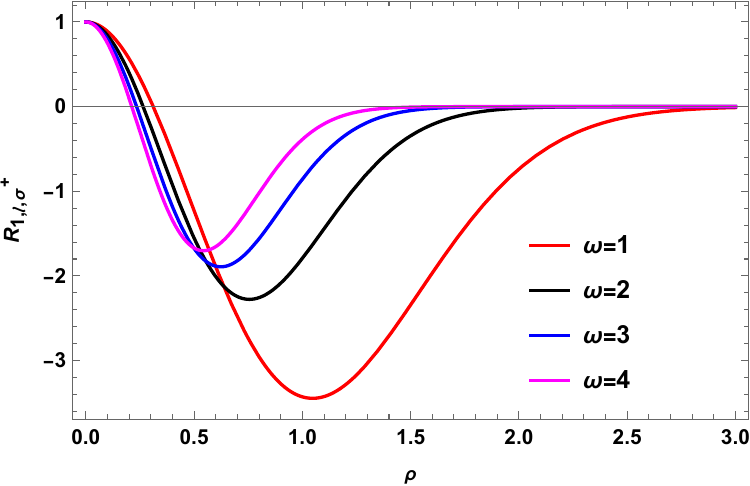}}\quad
\subfloat[$\alpha=0.5$]{\centering{}\includegraphics[scale=0.32]{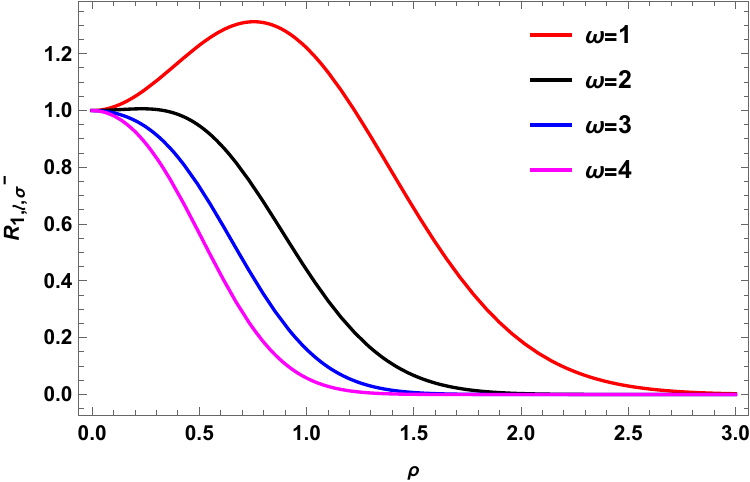}}
\centering{}\caption{The wave function $R_{1,\ell,\sigma}$ of Eq. (\ref{ee12}) for $\ell=1$-state. Here, $R^{+}$ indicates positive sign within the expression and $R^{-}$ that for negative sign.  Here $M=1$, $\kappa=0.5$, $\sigma=0.1$.}\label{fig-wave-function7}
\end{figure}
\par\end{center}

Equation (\ref{ee11}) describes the ground state energy level of a harmonic oscillator, while Equation (\ref{ee12}) provides the corresponding radial wave function for the radial mode defined by $n=1$. We have generated Figure 10 for the $\ell=0$ state to show the behavior of the ground state energy level with respect to $\alpha$ and $\omega$. In Figures 10(a) and 10(b), we observe that the energy level gradually increases. In Figures 10(c) and 10(d), the energy level increases linearly. We have also plotted the ground state energy level for the $\ell=1$ state in Figure 11, demonstrating its behavior with different parameters. Similarly, we have generated Figure 12 for the $\ell=0$ state and Figure 13 for the $\ell=1$ state, illustrating the behavior of the ground state radial wave function. These figures show how the radial wave function changes with different parameter values.

\section{Conclusions}

The significance of the eigenvalue solutions obtained in the context of curved space is much greater compared to those in flat space. This is because gravitational effects are taken into account when investigating the quantum dynamics of particles, leading to modified results. Numerous studies have explored the quantum motions of spin-0 and spin-1/2 particles in various curved and conical geometries. In this paper, we aimed to study the quantum dynamics of a harmonic oscillator system within the framework of topologically charged Eddington-inspired Born-Infeld (EiBI) gravity space-time, in the presence of a Wu-Yang magnetic monopole (WYMM). We analyzed how these modified theories alter the quantum dynamics of a harmonic oscillator system.

In Section 2, we investigated the quantum dynamics of a harmonic oscillator in a topologically charged EiBI-gravity space-time. We derived the radial equation and obtained an analytical solution, presenting the ground state energy level $E_{1,\ell}$ and the corresponding wave function $R_{1,\ell}$. various parameters, such as GM parameter $\alpha$, the strength of WYMM $\sigma$, the Born-Infeld parameter $\kappa$, and the oscillator frequency $\omega$, alters the eigenvalue solutions of the harmonic oscillator. It is worth noting that, unlike the results in Ref. \cite{EAFB}, we could not derive a closed-form expression for the energy spectrum. To illustrate the nature of the energy levels and the radial wave function, we generated several figures showing their behavior as various parameters change.

In Section 3, we studied the quantum dynamics of a harmonic oscillator in GM space-time without EiBI gravity, i.e., with the EiBI parameter $\kappa=0$. Additionally, we incorporated a WYMM into the quantum system and derived the radial equation in the presence of an external potential $V({\bf r})$. We then solved this radial equation under two scenarios: zero potential (Sub-section 3.1) and an inverse square potential (Sub-section 3.2). In each sub-section, we presented analytical solutions and obtained closed-form expressions for the energy spectrum and the radial wave functions. Our analysis showed that various factors, such as the GM parameter $\alpha$, the strength of the WYMM characterized by $\sigma$, the potential depth characterized by $\eta$, and the oscillator frequency $\omega$, significantly alter the eigenvalue solutions of the harmonic oscillator, leading to deviations from the flat space results. To illustrate the behavior of the energy spectrum and the radial wave function, we generated several figures showing the effects of varying these parameters.

In Section 4, we investigated the quantum dynamics of a harmonic oscillator in a topologically charged EiBI-gravity space-time in the presence of a Wu-Yang magnetic monopole (WYMM) field. Following a similar procedure as in Section 2, we presented the ground state energy level and the corresponding wave function of the harmonic oscillator. Our study showed that various parameters, such as the GM parameter $\alpha$, the strength of the WYMM $\sigma$, the Born-Infeld parameter $\kappa$, and the oscillator frequency $\omega$, significantly alter the eigenvalue solutions of the harmonic oscillator.

Our study of the quantum dynamics of a harmonic oscillator in the context of topologically charged Born-Infeld gravity space-time sheds light on the complex interactions between the quantum mechanical systems and gravitational fields. We have demonstrated how the Eddington parameter influences the behavior of the harmonic oscillator in a global monopole space-time. Our results underscore the importance of considering nonlinear gravitational effects in high-energy quantum systems, showing that the Eddington parameter plays a crucial role in defining corrections to the energy levels and wave-functions.

Additionally, we explored the behavior of the harmonic oscillator in the absence of EiBI gravity effects ($\kappa=0$), focusing on the presence of a WYMM and an inverse square potential. We found that the strength of the WYMM field and the potential depth significantly alter the harmonic oscillator's behavior in a global monopole space-time in addition to other parameters. Unlike the conventional quantum harmonic oscillator, the inclusion of an inverse square potential introduces additional complexity to the system, resulting in a richer structure of the energy spectrum and wave-functions.

\section*{Data Availability}

No data were generated or analysed in this study.

\section*{Conflict of Interests}

There is no conflicts of interests.

\section*{Funding Statement}

No fund has received for this paper.

\section*{Acknowledgements}

F.A. acknowledges the Inter University Centre for Astronomy and Astrophysics (IUCAA), Pune, India for granting visiting associateship.

\section*{Appendix: The Confluent Heun Equation}\label{appA}

\setcounter{equation}{0}
\renewcommand{\theequation}{A.\arabic{equation}}

The standard form of the confluent Heun equation is Refs. \cite{AR, SYS}
\begin{eqnarray}
H''(x)+\Big[\Xi+\frac{\beta+1}{x}+\frac{\gamma+1}{x-1}\Big]\,H'(x)+\Big[\frac{\mu}{x}+\frac{\nu}{x-1} \Big]\,H(x)=0,\label{A.1}
\end{eqnarray}
where $H(x)=H_{c}(\Xi, \beta, \gamma, \delta, \chi;x)$ is the confluent Heun function. The parameters $\mu$ and
$\nu$ given in the last term of Eq. (B.1) are defined as
\begin{eqnarray}
\mu=\frac{1}{2}\,(\Xi+\Xi\,\beta-\beta-\beta\,\gamma-\gamma)-\chi,\quad \nu=\frac{1}{2}\,(\Xi+\Xi\,\gamma+\beta+\beta\,\gamma+\gamma)+\delta+\chi.\label{A.2}
\end{eqnarray}
By using the Frobenius method Refs. \cite{GBA}, we can obtain a polynomial solution to the confluent Heun equation. Let us write the confluent Heun function as a power series around the origin,
\begin{equation}
H(x)=\sum^{\infty}_{i=0}\,d_{i}\,x^{i},\label{A.3}
\end{equation}
where $d_i$ are the coefficients. 

Thereby, substituting this power series in the Eq. (A.1) we obtain a few coefficients:
\begin{eqnarray}
d_1=-\frac{\mu}{\beta+1}\,d_0,\label{A.4}
\end{eqnarray}
with the following recurrence relation
\begin{eqnarray}
d_{k+2}=\frac{1}{(k+2)(k+2+\beta)}\Big[\Big\{(k+1)(k+2+\beta+\gamma-\Xi)-\mu\Big\}\,d_{k+1}+(\Xi\,k+\mu+\nu)\,d_{k}\Big].\label{A.5}
\end{eqnarray}

Therefore, from the Eq. (A.5), the confluent Heun series becomes a polynomial of degree $n$ when we impose two conditions:
\begin{eqnarray}
d_{n+1}=0,\quad \delta=-\Xi\,\Big[n+\frac{1}{2}\,(2+\beta+\gamma) \Big],\label{A.6}
\end{eqnarray}
where $n=1,2,3,....$. But, we do not know if a closed expression for the asymptotic behaviour of the confluent Heun function for large values of its argument exists.

\end{document}